Roadmap

# Roadmap for Unconventional Computing with Nanotechnology

Giovanni Finocchio[1,39,40], Jean Anne C. Incorvia[2], Joseph S. Friedman[3], Qu Yang[4], Anna Giordano[5], Julie Grollier[6], Hyunsoo Yang[4], Florin Ciubotaru[7], Andrii Chumak[8], Azad J. Naeemi[9], Sorin D. Cotofana[10], Riccardo Tomasello[11], Christos Panagopoulos[12], Mario Carpentieri[11], Peng Lin[13], Gang Pan[13], J. Joshua Yang[14], Aida Todri-Sanial[15], Gabriele Boschetto[16], Kremena Makasheva[17], Vinod K. Sangwan[18], Amit Ranjan Trivedi[19], Mark C. Hersam[18], Kerem Y. Camsari[20], Peter L. McMahon[21], Supriyo Datta[22], Belita Koiller[23], Gabriel H. Aguilar[23], Guilherme P. Temporão[24], Davi R. Rodrigues[11,40], Satoshi Sunada[25], Karin Everschor-Sitte[26], Kosuke Tatsumura[27], Hayato Goto[27], Vito Puliafito[11], Johan Åkerman[28], Hiroki Takesue[29], Massimiliano Di Ventra[30], Yuriy V. Pershin[31], Saibal Mukhopadhyay[9], Kaushik Roy[22], I-Ting Wang[32], Wang Kang[33], Yao Zhu[34], Brajesh Kumar Kaushik[35], Jennifer Hasler[9], Samiran Ganguly[36], Avik W. Ghosh[37], William Levy[37], Vwani Roychowdhury[38], Supriyo Bandyopadhyay[36,39,40]

[1] Department of Mathematical and Computer Sciences, Physical Sciences and Earth Sciences, University of Messina, 98166 Messina, Italy

[2] Chandra Family Department of Electrical and Computer Engineering, University of Texas at Austin, Austin, TX 78712, USA

[3] Department of Electrical and Computer Engineering, University of Texas at Dallas, Richardson, TX 75080, USA

[4] Department of Electrical and Computer Engineering, National University of Singapore, Singapore 117576

[5] Department of Engineering, University of Messina, 98166 Messina, Italy

[6] Laboratoire Albert Fert, CNRS, Thales, Université Paris-Saclay, 91767 Palaiseau, France

[7] IMEC, Leuven 3001, Belgium

[8] Faculty of Physics, University of Vienna, Vienna 1090, Austria

[9] School of Electrical & Computer Engineering, Georgia Institute of Technology, Atlanta, GA 30332, USA

[10] Quantum and Computer Engineering Department, Delft University of Technology, Delft, CD 2628, Netherlands

[11] Department of Electrical and Information Engineering, Politecnico di Bari, 70125 Bari, Italy

[12] Division of Physics and Applied Physics, School of Physical and Mathematical Sciences, Nanyang Technological University, Singapore, 639798, Singapore

[13] College of Computer Science and Technology, Zhejiang University, Hangzhou, 310013, China

[14] Department of Electrical and Computer Engineering, University of Southern California, Los Angeles, CA 90089, USA

[15] Electrical Engineering Department, Eindhoven University of Technology, Eindhoven, AZ 5612, Netherlands

[16] Microelectronics Department, LIRMM, Université de Montpellier, 34095 Montpellier, France

[17] Laplace, CNRS, UT3, INPT, University of Toulouse, 31062 Toulouse, France

[18] Department of Materials Science and Engineering, Northwestern University, Evanston, IL 60208, USA

[19] Department of Electrical and Computer Engineering, University of Illinois at Chicago, Chicago, IL 60607, USA

[20] Department of Electrical and Computer Engineering, University of California at Santa Barbara, Santa Barbara, CA 93106, USA

[21] School of Applied and Engineering Physics, Cornell University, Ithaca, NY 14850, USA

[22] Elmore Family School of Electrical and Computer Engineering, Purdue University, West Lafayette, IN 47907, USA

[23] Instituto de Física, Pontifícia Universidade Católica do Rio de Janeiro, 22451-900 Rio de Janeiro, RJ, Brasil

[24] Center for Telecommunications Studies, Pontifical Catholic University of Rio de Janeiro, 22451-900 Rio de Janeiro, RJ, Brazil

[25] Faculty of Mechanical Engineering, Institute of Science and Engineering, Kanazawa University, Kanazawa, Ishikawa 920-1192, Japan

[26] Faculty of Physics and Center for Nanointegration Duisburg-Essen (CENIDE), University of Duisburg-Essen, Duisburg, D 47057, Germany

[27] Toshiba Corporation, Kawasaki 212-8582, Japan

[28] Department of Physics, University of Gothenburg, 41296 Göteborg, Sweden

[29] NTT Corporation, Atsugi, Kanagawa, 243-0198 Japan

[30] Department of Physics, University of California, San Diego, La Jolla, CA 92093, USA

[31] Department of Physics and Astronomy, University of South Carolina, Columbia, South Carolina 29208, USA

[32] Taiwan Semiconductor Research Institute, Hsinchu 300091, Taiwan

[33] School of Integrated Circuit Science and Engineering, Beihang University, Beijing 100191, China

[34] Institute of Microelectronics, Singapore 117685, Singapore

[35] Department of Electronics and Communication Engineering, Indian Institute of Technology-Roorkee, Roorkee, Uttarakhand 247667, India

[36] Department of Electrical and Computer Engineering, Virginia Commonwealth University, Richmond, VA 23284, USA

[37] Department of Electrical and Computer Engineering, University of Virginia, Charlottesville, VA 22904, USA

[38] Department of Electrical and Computer Engineering, University of California at Los Angeles, CA 90095, USA

[39] Guest editors of the Roadmap.

[40] Author to whom any correspondence should be addressed.

E-mails: giovanni.finocchio@unime.it (Giovanni Finocchio), davi.rodrigues@poliba.it (Davi R. Rodrigues), sbandy@vcu.edu (Supriyo Bandyopadhyay)

**Abstract**

In the "Beyond Moore's Law" era, with increasing edge intelligence, domain-specific computing embracing unconventional approaches will become increasingly prevalent. At the same time, adopting a variety of nanotechnologies will offer benefits in energy cost, computational speed, reduced footprint, cyber resilience, and processing power. The time is ripe for a roadmap for unconventional computing with nanotechnologies to guide future research, and this collection aims to fill that need. The authors provide a comprehensive roadmap for neuromorphic computing using electron spins, memristive devices, two-dimensional nanomaterials, nanomagnets, and various dynamical systems. They also address other paradigms such as Ising machines, Bayesian inference engines, probabilistic computing with p-bits, processing in memory, quantum memories and algorithms, computing with skyrmions and spin waves, and brain-inspired computing for incremental learning and problem-solving in severely resource-constrained environments. These approaches have advantages over traditional Boolean computing based on von Neumann architecture. As the computational requirements for artificial intelligence grow 50 times faster than Moore's Law for electronics, more unconventional approaches to computing and signal processing will appear on the horizon, and this roadmap will help identify future needs and challenges. In a very fertile field, experts in the field aim to present some of the dominant and most promising technologies for unconventional computing that will be around for some time to come. Within a holistic approach, the goal is to provide pathways for solidifying the field and guiding future impactful discoveries.

**Contents**

# Introduction

In the "Beyond Moore's Law" era, domain-specific computing embracing unconventional approaches will become increasingly prevalent with increasing edge intelligence. Several shortcomings plague current computational models and the underlying technologies: (1) exorbitant energy cost, which drains resources and will ultimately affect our environment; (2) slowing of Dennard scaling and Moore's law since we are reaching the limits of transistor downscaling, which would impact the economy in the long run; (3) difficulty in reconciling fast switching times with long memory retention: conventional computing systems face a trade-off between memory speed and capacity owing to hierarchical memory structures, which would have adverse consequences for a world that is always data-hungry; (4) limited parallelism: while conventional computing can achieve some parallelism through multi-core processors, not all computing tasks can be efficiently parallelized, leading to slow data processing; and (5) the von Neumann bottleneck which limits the size, weight and power (SWaP) of our current computational technologies.

New unconventional computing paradigms have emerged to contend with some or all of these challenges. At the same time, adopting a wide variety of nanotechnologies to implement unconventional computing will benefit energy cost, computational speed, reduced footprint, cyber-resilience, and data processing prowess. The time is ripe to lay out a roadmap for unconventional computing with nanotechnologies to guide future research, and this collection aims to fulfill that need. The authors provide a holistic roadmap for unconventional computing with electron spins, memristive devices, nanomaterials, mixed-dimensional heterojunctions, nanomagnets, and assorted dynamical systems. The authors address the similarities and differences between the paradigms discussed in the manuscript, emphasizing the underlying connections. For example, the manuscript distinguishes between neuromorphic computing and brain-inspired computing. Although both are based on the behavior of a biological brain, the former is usually associated with neural networks, while the latter aims to delve deeper into the architecture of the brain, taking into account regions with different memory and processing functions. They also address other paradigms for solving combinatorial optimization and graph-theoretic problems, such as Ising machines and simulated bifurcation, computing in the presence of uncertainties such as Bayesian inference engines, probabilistic computing with p-bits, processing in memory to circumvent the von-Neumann bottleneck, quantum memories and algorithms for solving NP problems, computing with skyrmions and spin waves for massive parallelism, and brain-inspired computing for incremental learning and solving problems in severely environments with severe constraints on resources, see Figure. All of these approaches have advantages over conventional Boolean computing predicated on the von-Neumann architecture. With the computational need for artificial intelligence growing at a rate 50-fold faster than Moore's law for electronics, more unconventional approaches to computing and signal processing will appear on the horizon, and this roadmap will aid in identifying future needs and challenges.

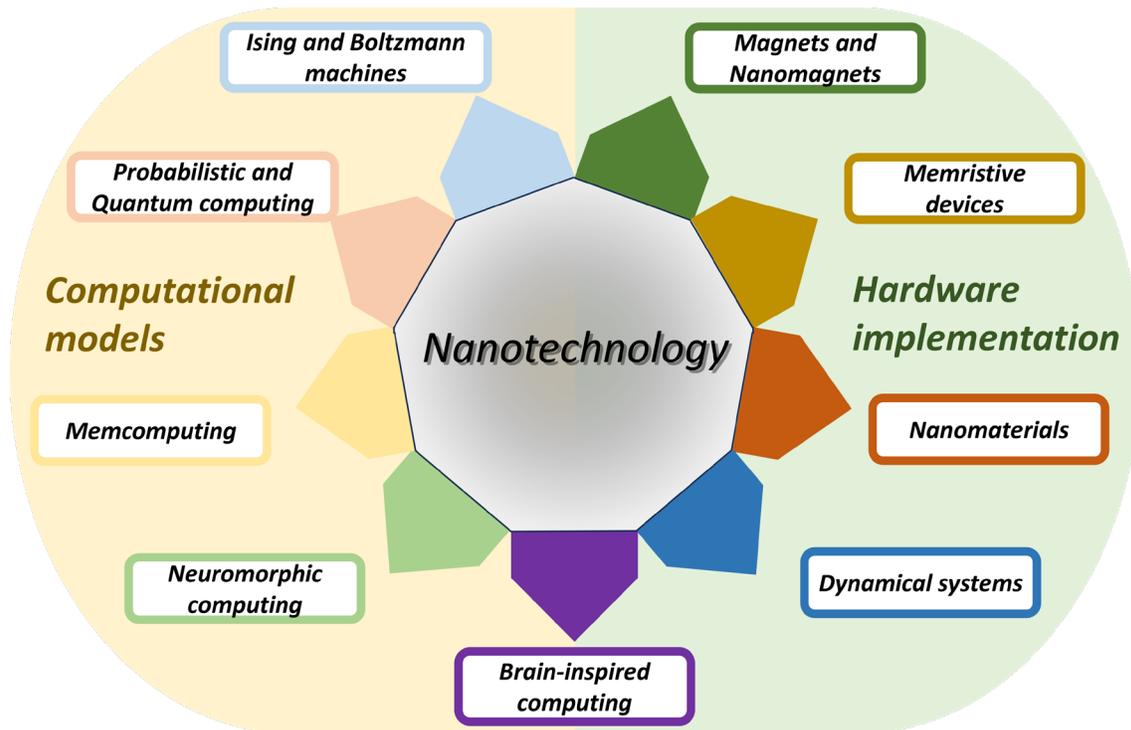

**Figure.** Illustrate the potential applications of nanotechnology. This roadmap explores hardware implementations across a spectrum of computational paradigms and highlights computational models that are poised to take full advantage of nanotechnology components. We will focus on cutting-edge computational models and hardware implementations, describing their current status and challenges, as well as current nanotechnology efforts to address these challenges. Section 1 covers hardware implementation using magnets and nanomagnets for neuromorphic computing. Section 2 covers memristive devicesand their potential use for unconventional computing.. Section 3 covers the use of nanomaterials for unconventional computing. Section 4 covers probabilistic and quantum computing. Section 5 considers the use physical systems and physics-inspired models, such as Ising and Boltzmann machines as well as memcomputing, for unconventional computing. Section 6 covers in-memory computing. Section 7 covers brain-inspired computing.

In particular, Ising and Boltzmann machines are similar in their (physics-based) computing approaches. In contrast, simulated bifurcation is related to both and can be viewed as a natural progression of the underlying idea. They embed the solution to a problem in the ground state configuration of several interacting devices, which are programmed by biasing each one and adjusting the weight of the synaptic connection between them. They have two significant advantages. First, since they compute by relaxing to the ground state, no external energy needs to be pumped into them to maintain them in an excited state. That makes them highly frugal in the use of energy. Second, they are very forgiving of errors since the computational activity is elicited from the cooperative actions of many devices working in unison, and the failure of one or few devices does not impair circuit functionality. Computing with large-scale dynamical systems, whose natural response mimics the execution of algorithms without any software, shares these characteristics. For example, the combination of dynamical systems with the memcomputing paradigm is a promising direction for solving combinatorial optimization problems, including prime factorization, at scale. Brain-inspired computing can also become more efficient with this feature. Probabilistic computing employing p-bits and quantum computing utilizing qubits also have some mutual connection. The p-bit is sometimes referred to as a poor person's qubit. The difference is that a p-bit is sometimes the bit 0 and sometimes the bit 1 (with tailored probabilities) but never simultaneously both 0 and 1, while a qubit is simultaneously both 0 and 1 all the time but collapses to either bit with different (tailored) probabilities when a measurement is made. Both probabilistic and quantum computing have been shown to be capable of solving combinatorial optimization problems such as the traveling salesman problem and maximum satisfiability, although a thorough scalability analysis is needed. In the same vein, there is a similarity between computing with skyrmions and spin waves, not just because they are both magnetic entities but because both can be adapted to either analog or digital information processing. Skyrmions may also have a role to play in quantum computing.

Furthermore, memristive systems promise the realization of ultrafast and ultradense memory elements that can mimic a wide range of functions ranging from matrix multiplication to bio-realistic synapses for hardware accelerators and neural networks, respectively. In particular, emerging memtransistors and mixed-dimensional heterojunctions can realize bio-realistic neuronal functions such as input-adaptive learning, continuous learning, heterogeneous plasticity, and complex spiking behavior that can simplify circuit architectures. Nanomaterials enable printed neuromorphic devices to integrate with bio-compatible sensors and flexible electronics applications. While these nascent ideas face different challenges than more mature technologies, such as floating gate memories, they also promise alternative paradigms that can circumvent technological hurdles due to inherently superior physical properties, form factors, and device metrics. One of their primary advantages is the tunability of response functions that establish a fertile ground to realize devices and circuits for the circuit realization of cortical architectures and processes.

This article not only addresses paradigms but also addresses technologies that are best suited to a given approach. Most algorithms can be executed with different devices and hardware, whether nanomagnets, nano-CMOS, memristors, or something else. In other words, they are generally hardware agnostic. However, some algorithms run best on a particular class of devices, depending on their specific properties, such as non-volatility, intrinsic stochasticity, time nonlocality, and memory. Artificial neural networks adapt well to both memristors and magnetic devices, as well as to CMOS. Addressing and updating large memristor crossbar arrays is a challenge where dual-gated memtransistors based on two-dimensional materials can simplify the architecture and operation of these synaptic circuits. Probabilistic computing's ideal hardware will be low-barrier nanomagnets, perhaps also CMOS. Ising machines and others that rely on the relaxation of an interacting assembly of devices to their many-body ground state can employ various technologies. However, some may be better than others depending on the type of problem being solved. One technology receiving widespread attention involves coupled oscillators involving spin Hall nano-oscillators or CMOS. In this article, each paradigm is illustrated with a particular technological substrate, whether CMOS, nanomagnets, memristive elements, or something else, because it is easy and convenient to implement the paradigm with that technology. Additionally, the chosen technology may reduce SWaP, which is an important consideration.

The manuscript is organized as follows: Section 1 deals with using magnetism to implement unconventional computing at the nanoscale. Sections 1.1 and 1.2 focus on spintronic technology based on magnetic tunnel junctions, while sections 1.3 and 1.4 discuss spin-wave and skyrmion-based approaches. Section 2 discusses the use of memresistors for unconventional computing. Section 3 discusses nanomaterial systems for hardware implementation of unconventional computing, with Section 3.2 focusing on two-dimensional materials. Section 4 delves into the realm of using probabilities for complex computations, which includes the discussion of probabilistic computing in Section 4.1 and quantum computing in Section 4.2. Section 5 discusses the use of dynamical systems for complex computations. Section 5.1 discusses how unconventional computation can exploit the intrinsic complex dynamics of physical systems. Sections 5.2 and 5.3 consider models and hardware implementations for Boltzman and Ising machines. Section 5.4 deals with memcomputing. Section 6 deals with in-memory computing. Finally, Section 7 considers brain-inspired computing, taking into account the complex interactions between the hippocampus and the cortex.

Each technology, of course, comes with its challenges. The primary obstacles in spintronic and magnetic technologies are device-to-device variations, sensitivity to defects, and the deleterious effects of thermal noise. In the case of spin wave devices, the challenge is to find efficient interfaces for input and output. With skyrmionic computing, the roadblocks involve inadequate control over skyrmions' size, stability, creation, annihilation, and electrical readout. Memristive technologies may face some material challenges and may have to contend with device-to-device variability and the experienced difficulty of large-scale integration. Scalability is also an issue. The advances on the nanomaterial level are indispensable for the transition to the system level of application. Emerging tunable memristive systems such as memtransistors from two-dimensional materials and mixed-dimensional heterojunctions possess additional challenges, including wafer-scale growth, transfer, and possible integration as back-end-of-the-line (BEOL) circuits. For probabilistic computing, the main challenges are device-to-device variation, which hinders large-scale integration, and slow computational speed when implemented with low barrier nanomagnets made of common ferromagnets owing to the small flips per second determining the p-computer rate. Control, coherence, and readout are the looming obstacles for spin-based qubits. The need for efficient simulated annealing techniques to avoid metastable states (sub-optimal solutions) and to speed up computation challenges coupled oscillator technology for Ising machines. Compute-in-memory faces materials-related challenges and the complexity of implementation. Finally, brain-inspired computing has to contend with many daunting challenges, such as faithful reproduction of synaptogenesis and dendritogenesis, that are difficult to implement with artificial devices and circuits.

By its nature, no roadmap article can claim to be completely comprehensive, and we make no such claim either. New ideas and technologies germinate fast, and today's state of the art becomes obsolete tomorrow. Here, we have presented some of the dominant and most promising technologies for unconventional computing that we believe will endure for some time.

Finally, a critical need is to create standardizing metrics to benchmark different approaches to unconventional computing. Their performances should be evaluated at various levels, including hardware (device and circuit) and algorithmic. Developing standardizing metrics is a complex task that requires interdisciplinary collaboration and a deep understanding of the underlying principles and design goals. However, by establishing rigorous metrics, researchers and practitioners can effectively compare and evaluate the performance of different unconventional computing systems. Via the applied holistic approach, this roadmap article could contribute to achieving that objective.

# 1.1 – Magnetic Architectures for Unconventional Computing

Jean Anne C. Incorvia, Chandra Family Department of Electrical and Computer Engineering, The University of Texas at Austin, Austin, TX 78712 (incorvia@austin.utexas.edu)

Supriyo Bandyopadhyay, Department of Electrical and Computer Engineering, Virginia Commonwealth University, Richmond, VA 23284 (sbandy@vcu.edu)

Joseph S. Friedman, Department of Electrical and Computer Engineering, The University of Texas at Dallas, Richardson, TX 75080 (Joseph.Friedman@utdallas.edu)

**Status**

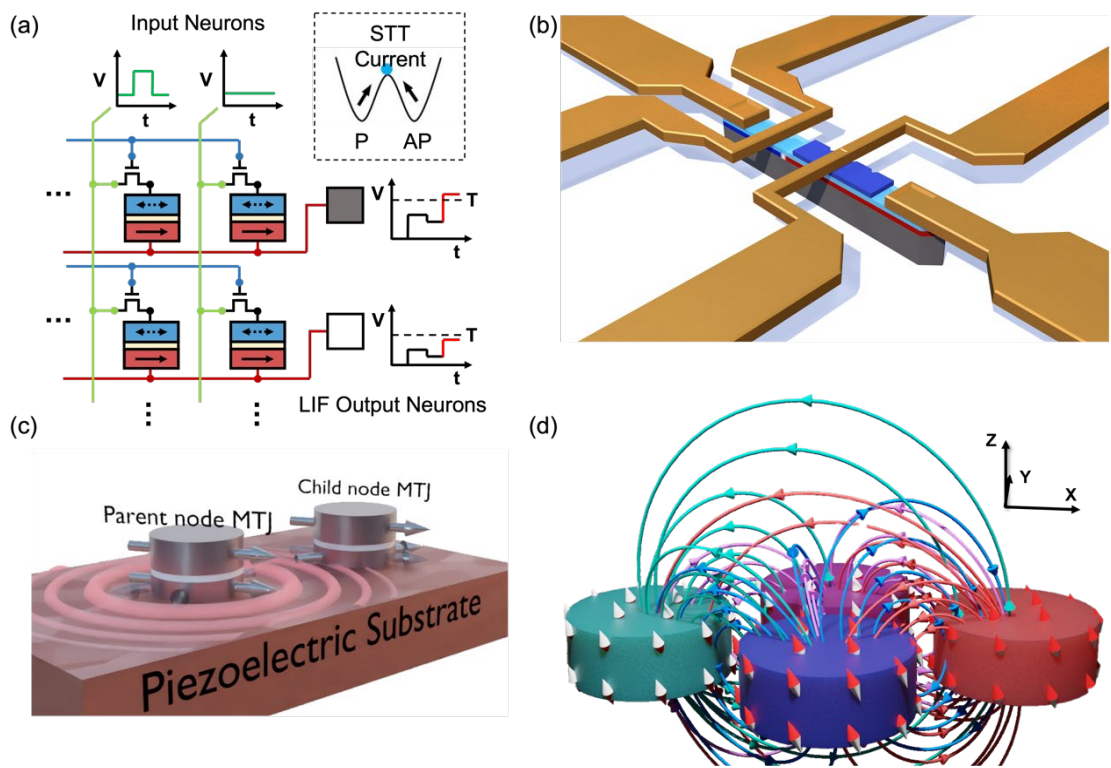

**Figure 1.** (a) Neuromorphic crossbar array with the stochastic writing of MTJ synapses. (b) Example domain wall-MTJ, here patterned as a synapse, with patterned blue/white/red domain wall track and blue output tunnel junction, including notches to control the domain wall position. (c) A two-node Bayesian network implemented with two MTJs. (d) Reservoir computer comprised of frustrated nanomagnets.

Next-generation unconventional computing will address key needs and problems for processing increasingly large and unstructured data workloads, as well as the increase in edge computing devices and corresponding energy constraints. Some problems that unconventional computing addresses include the bottleneck between compute and memory; the large energy and delay penalty of analog-to-digital conversion; computing with small energy budgets; and application-specific computing with balanced energy, time, and precision needs, since precise computing is not always required.

Magnetic thin films, both continuous and patterned into nanomagnets, have a long history in computing, starting with hard disk drives and including today's spin transfer torque and spin orbit torque-based magnetic random access memory (STT-MRAM, SOT-MRAM). Unconventional computing hardware (neuromorphic, Bayesian, Boltzmann machines) implemented with magnetic devices, e.g., magnetic tunnel junctions (MTJs), are attractive since the constituent elements are non-volatile and could be extremely energy-efficient. The device characteristics and inter-device interactions, which depend on the energy barrier within the free layer of the MTJ, can be tailored and controlled through multiple simultaneous knobs, such as current, voltage, strain, magnetic fields, and by both DC and AC inputs. This offers immense flexibility in designing hardware accelerators for machine learning such as binary stochastic neurons (BSNs), neuromorphic components like synapses, Ising machines, etc.

The MTJ and corresponding devices also benefit from high endurance in switching the magnetic state, and from the fact that, under normal operation, the resistance states can be set in a controllable way, without drift over time or over cycles. This stability and robustness of the bit state control (not necessarily the states themselves, which can be tuned between stable and stochastic) can compensate for some of the challenges MTJs face.

Whereas MTJs with low energy barriers exhibit constant stochastic switching between resistance states, useful for BSNs, MTJ memory devices with high energy barriers exhibit an alternative stochastic phenomenon: the switching between the two stable states is intrinsically stochastic. This stochastic writing process provides analog behavior to these binary memory devices, enabling their use in neuromorphic systems of the type described in Fig. 1(a). Furthermore, the binary MTJ states are inherently robust against the variations and stochastic behavior that plagues memristors and phase-change memory, thereby making non-volatile MTJ synapses a promising technology for neuromorphic computing.

Neural network crossbar arrays can be implemented using the nanomagnet as both the artificial synapse and the artificial neuron. By using a top-pinned MTJ stack and extending the bottom magnetic layer into a longer track, the MTJ can be configured as a domain wall-magnetic tunnel junction (DW-MTJ), shown in Fig. 1(b). Subsequent choice of patterning can then have the device show analog resistance states as a synapse [1] or as a neuron [2]. While a domain wall, or similarly a magnetic skyrmion, can be harder to control than a single-domain nanomagnet, it provides additional bio-mimetic functions for unconventional computing such as time delays, stochastic pinning and depinning, and frequency-based switching. It can also benefit from magnetic field interactions between the domain walls of the devices.

Belief networks (Bayesian inference engines) are another genre of unconventional computers for computing in the presence of uncertainty. They are difficult to implement with most technologies since they require *non-reciprocal* synapses. Simple 2-node networks consist of a parent and a child node where the child node's state is correlated with that of the parent, but not the other way around. Two dipole-coupled MTJs of different shapes built on a piezoelectric substrate can implement this paradigm easily, show in Fig. 1(c). The degree of correlation or anti-correlation between the nodes can be varied with global strain applied to both MTJs via the piezoelectric [3] and this can enable Bayesian inference [4]. The synaptic connection between the nodes is dipole coupling, which consumes no area on the chip and dissipates no energy since it does not involve current flow.

While trained neuromorphic computing systems promise exceptional capabilities, the training process incurs significant hardware costs in terms of energy, area, and speed. Reservoir computing therefore provides an opportunity to avoid those costs by using a system that requires minimal training. In particular, the bulk of the system is untrained, while only a single output layer must be trained. Nanomagnetism naturally provides such reservoirs, as irregular arrays of closely-packed nanomagnets exhibit frustration that produces complex physical dynamics and hysteresis, shown in Fig. 1(d). All these extra-ordinary capabilities make magnetic architectures for unconventional computing unique and attractive.

**Current and Future Challenges**

The year 2021 heralded the first three experimental demonstrations of neural networks with synapse weights encoded in binary MTJ states; all three performed some type of recognition task. In the simplest of these experiments, a 4x2 single-layer neuromorphic network was directly implemented with MTJ synapses [5] to perform vector-matrix multiplication. More complex MTJ-based synapse structures were used in a two-layer (13×6 + 6×3) network as well as a 64x64 single-layer network [6]. The key future challenges for this neuromorphic computing approach are scaling to large network dimensions and the experimental demonstration of learning through stochastic switching.

DW-MTJs also have been demonstrated recently [1, 2, 7]. Clear needs are better understanding and control of the domain wall behavior over many cycles, especially without needing to refresh the devices; all-electrical control without the need of external magnetic fields to aid domain wall movement; scaling down to modern feature sizes; and scaling up for larger circuit demonstrations, including better understanding of device-to-device variations and their impact on the unconventional computing applications.

As a first step towards the development of reservoir computers based on frustrated nanomagnetism, micromagnetic simulation studies have demonstrated their memory capacity and expressivity [8]. These systems have been shown to successfully perform complex classification tasks, including waveform identification, Boolean operations based on previous inputs, and observation and prediction of dynamical discrete time series. Furthermore, comparative simulation studies indicate a 60x improvement in energy efficiency relative to conventional CMOS systems [8]. However, experimental demonstration and proof of concept remain a significant challenge.

Neuromorphic computing is generally much more forgiving of switching errors than Boolean logic, but it is not necessarily very tolerant of large device-to-device variations. The response time of BSNs, for example, can change dramatically in the presence of fabrication defects or slight shape variations, which results in significant device-to-device variations that is a challenge for large scale networks. One way to counter this is to adopt hardware aware in-situ learning [9]. Another is to replace common ferromagnets used in MTJs with dilute magnetic semiconductors which have several orders of magnitude lower saturation magnetization. That makes the energy barriers in the nanomagnets much less sensitive to shape and size variations and suppresses device-to-device variations [10].

A challenge with ferromagnetic devices is the relatively slow switching speed of ~1 ns which creates a bottleneck in training and inference in both recurrent and deep neural networks. There has been some recent interest in harnessing anti-ferromagnetic materials for synapses and they are capable of much higher speed. This is a nascent field, but important discoveries may be around the corner. All these challenges make magnetic architectures for unconventional computing a fertile field of research.

**Advances in Science and Technology to Meet Challenges**

One of the critical challenges for neural networks based on both conventional MTJs and DW-MTJs is efficient network training. As conventional supervised learning algorithms (e.g., backpropagation) become increasingly complex when scaled to large and deep networks, the hardware costs for implementing this mathematical circuitry hinder the development of neural networks with online learning. Preliminary explorations of unsupervised learning algorithms with MTJs and DW-MTJs [5] indicate that local Hebbian learning rules can be used with feedback circuits to efficiently train neural networks with minimal energy, speed, and area costs.

Realization of reservoir computing systems based on frustrated nanomagnets will require advances in experimental techniques for providing the input signals while contemporaneously measuring the magnetization of the various nanomagnets that make up the reservoir. The inputs can be provided via STT switching, and the output magnetizations can be read through an MTJ (preliminary experimental efforts may focus on imaging the output). These output signals must be fed to a single trained layer.

Domain wall creep and the stochasticity of domain wall motion at room temperature pose significant obstacles to DW-MTJ technologies. Recent progress with notched structures [1,2] have ameliorated some of the difficulties, but further material research is needed to find possibly simpler solutions where the intrinsic material properties may be able to suppress or control the stochastic behaviour of domain wall and skrymion motion.

A well-known challenge with magnetic devices such as MTJs is the low on/off ratio and low overall resistance, which typically results in low training accuracies in neuromorphic architectures [6]. Research is needed to find proper material combinations to increase the tunnelling magneto-resistance or on/off ratios of MTJs to alleviate this problem. This field has a long history and unfortunately has been slow in making progress. However, its importance cannot be overstated since a high on/off ratio provides a wider range of synaptic weights and improves error tolerance.

Challenges in patterning MTJ-based structures leads to device-to-device variation that can increase with reducing feature size. This challenge compounds with the low on/off ratio to blur the difference between the 0 and 1 state, and is even more of an issue if more resistance levels are desired between the 0 and 1. While good control of the device resistance states can help with this issue, better patterning methods are needed, as well as more effort in circuit design to design around these challenges.

**Concluding Remarks**

The human brain consumes 1-100 fJ of energy per synaptic event. Magnetic devices can rival (or even eclipse) this energy efficiency. Their non-volatility offers additional architectural advantages, e.g., in reservoir computing [8].

The low energy consumption has other benefits: it provides hardware security, which is very important for artificial intelligence. Because of the low power requirement, these architectures can be embedded in edge devices that have minimal contact with the cloud and are therefore somewhat insulated from cloud-borne attacks. Additionally, they are inherently resilient against malign hardware. A hardware Trojan, no matter how surreptitious, will consume some energy and that can become comparable to (or exceed) the energy consumed by magnetic hardware. Therefore, Trojans can be easily detected with side channel monitoring.

Finally, neuromorphic computing with anti-ferromagnetic devices is a burgeoning area of research laden with promise and it can spawn new devices and architectures that will speed up training and inference tasks immensely. This is an exciting area of research that is about to bear fruit.

**Acknowledgements**

The work of S. B. in this field has been supported by the National Science Foundation under grants CCF-2001255 and CCF-2006843. The work of J. A. C. I. in this field has been supported by the National Science Foundation under grants EPMD-2225744, CCF-1910997, and CCF-2006753, as well as Sandia National Laboratories. The work of J. S. F. in this field has been supported by the National Science Foundation grants CCF-1910800 and CCF-2146439, and the Semiconductor Research Corporation.

## 1.2 – The impact of spintronics in neuromorphic computing

Qu Yang, Department of Electrical and Computer Engineering, National University of Singapore, Singapore 117576 (eleyaq@nus.edu.sg)
Anna Giordano, Department of Engineering, University of Messina, 98166 Messina, Italy (agiordano@unime.it)
Julie Grollier, Laboratoire Albert Fert, CNRS, Thales, Université Paris Saclay, 91767, Palaiseau, France (julie.grollier@cnrs-thales.fr)
Giovanni Finocchio, Department of Mathematical and Computer Sciences, Physical Sciences and Earth Sciences, University of Messina, 98166 Messina, Italy (gfinocchio@unime.it)
Hyunsoo Yang, Department of Electrical and Computer Engineering, National University of Singapore, Singapore 117576 (eleyang@nus.edu.sg)

**Status**

Neuromorphic spintronics aims to develop spintronic hardware devices and circuits with brain-inspired principles [1]. The conventional complementary metal–oxide–semiconductor (CMOS) neuron and synapse designs require numerous transistors and feedback mechanisms and would be unsuitable for developing modern artificial intelligence systems. Spintronics is a promising approach to neuromorphic computing as it potentially enables energy and area-efficient embedded applications by mimicking key features of biological synapses and neurons with a single device instead of using multiple electronic components [1, 2].

As discussed in Section 1.1, the main building block for neuromorphic spintronics is the magnetic tunnel junction (MTJ), which exhibits several unique characteristics over other technologies, including CMOS compatibility, low power consumption, outstanding read/write endurance, non-volatility, and fast speed [1]. Krysteczko et al. carried out the first work on the spintronic implementation of memristive functionalities by voltage-induced switching in MTJs [3]. Later on, different MTJ-based spintronic structures have been proposed to potentially offer solutions to neuronal computations with bio-fidelity [4]. For example, MTJs have been used for the realization of memristors for storing synaptic weights, activation functions of a neuron (such as ReLU-like and sigmoidal), reservoir computing and in-memory computing.

Fig. 1(a) illustrates an example of MTJ-based integrate-and-fire spiking neuron [4]. In this case, the current-induced spin orbit torque (SOT) integrates the domain wall motion (DWM). When the domain wall reaches the critical position (threshold), the neuron device spikes a "fire" signal. With similar principles, more sophisticated spin-based neuron models have been further developed, and similar structures with magnetic skyrmions (discussed in Section 1.4) instead of DWM as information carrier have also been constructed, see Fig. 1(b)-(c) [2, 5]. Fig. 1(d) presents a spin-torque nano-oscillator (STNO)-based neuron that emulates a Hodgkin-Huxley analogue model, surpassing the limitations of the integrate-and-fire model [6]. The proposed device harnesses the combined effects of magnetization dynamics and temperature fluctuations within the STNO, enabling the generation of a sequence of spikes whose frequency depends on the amplitude of the constant applied current. Moreover, various SOT neuromorphic solutions also have been demonstrated experimentally for the realization of ultrafast neuromorphic spintronics, field-free artificial neuron, auto-reset stochastic neuron, and stochastic artificial synapse.

Spintronics could be powerful in the development of neuromorphic computing because it enables the data processing and storage at a very local level. To this end, there have been a rich variety of spintronic materials and device designs for proof-of-concept neuromorphic computing implementations. Regarding the neural networks, the paradigm is now shifting from frame-based to event-based exploiting the idea of spiking neurons in spiking neural networks, an approach that is closer to the brain working principle. These novel research progresses have further aroused a research enthusiasm towards developing large-scale brain-inspired spintronic systems.

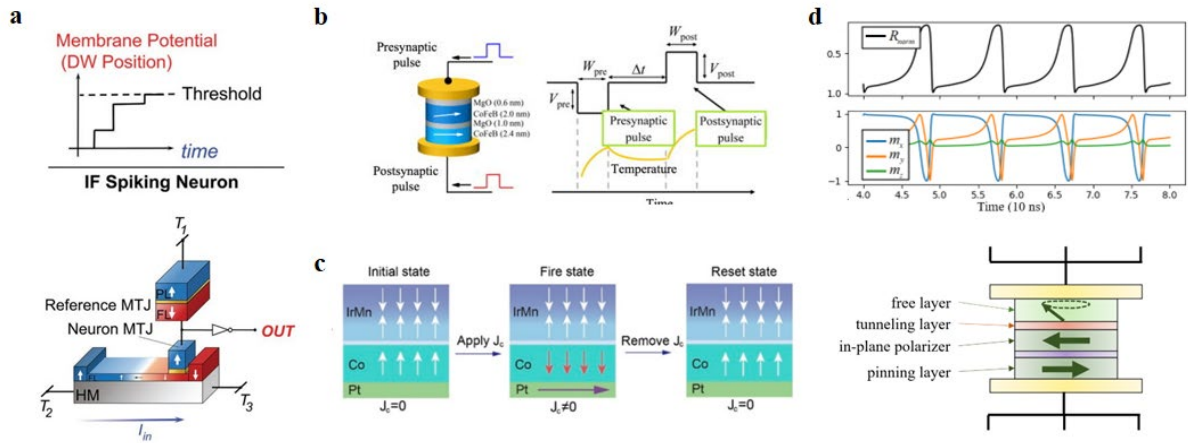

**Figure 1.** Different schemes for spintronics implementation of firing neurons. (a), (b) and (c) emulate the Leaky-Integrate, Fire and Reset model while (d) emulate the biorealistic Huxley-Hodgkin model. Figures reprinted with permission from [2], [4], [5], [6]. Copyright 2017, American Institute of Physics (AIP). Copyright 2022, American Institute of Physics (AIP). Copyright 2022, American Chemical Society (ACS), Copyright 2023, American Physical Society (APS)

**Current and Future Challenges**
There are important challenges to be overcome for further development of neuromorphic spintronics. One of the biggest challenges is that the read-out signal of spintronic approaches is quite small, making it difficult to read quickly. As discussed in Section 1.1, significant research endeavors have been dedicated to enhancing the tunneling magnetoresistance (TMR) in MTJs. Nonetheless, the resistance changes of MTJs (typically one to three on/off ratios) remain relatively modest compared to other memory technologies. To further address this issue, researchers have explored the integration of MTJs with CMOS technology to achieve a higher on/off ratio and lower leakage current [7].

Regarding spintronic neurons, typically a reset pulse with a sufficiently high magnitude (equal to or several times larger than that required for writing) and of opposite polarity is necessary [2]. This not only increases energy consumption and complexity of the chip, but also lowers the areal density for peripheral circuits required. Besides, an extra resetting step will decrease the operational speed of the neural circuit. The neural device will not be usable till it has been reset by a reset-pulse. Therefore, a bio-realistic neural device with the auto-reset functionality is desirable for energy-efficient and densely packed artificial neural networks.

Moreover, implementing spintronic hardware in neural networks has challenges in coupling control of each neuron. Synchronization of device properties instead of changing their individual response would be one promising way to extend spintronic approaches to multilayer neural networks [8]. As detailed in Section 1.1, addressing device variability, response speed, and circuit design challenges is crucial when connecting each neuron to potentially thousands of synapses in a neural network algorithm.

**Advances in Science and Technology to Meet Challenges**
Research efforts have been put to address the challenges encountered in spintronic neuromorphic computing. The spintronic memristor has been developed to emulate synaptic behaviors (Section 2.1). The inherent stochasticity in stochastic MTJ (S-MTJ) enables highly energy-efficient probabilistic computing tasks such as stochastic number generation and probabilistic spin logic operation (Section 4.1). Besides, the switching probability of S-MTJ is adjustable by an applied electric bias, and thus the junction can be utilized for the emulation of the Poisson neuron which generates a spiking train with a tunable firing rate.

Moreover, the unique features of STNO have been utilized for neuromorphic computing. Based on time multiplexing, the single oscillator can function as a reservoir computer, which is a special type of neural network for time series analysis [9]. In addition, the high tunability of STNOs facilitates the coupling with other oscillator devices and could emulate the synchronization of neurons. This is important for information sharing and processing. The classification of vowels at microwave frequencies has been experimentally demonstrated through the synchronization of STNOs [8]. The response of spin-diodes has also be used to mimic neurons in the non-linear regime, and synapses in the linear regime. Frequency multiplexing appears as a possible solution to build multilayer neural networks with STNO neurons and spin diodes synapses. Future research efforts should focus on the integration of spintronic devices (e.g. SOT devices) in the MTJ-based magnetic random-access memory (MRAM) architecture to increase the read-out signal via TMR. For the circuit design, the shared write channel-based SOT architecture can be explored to reduce the transistor count for large scaling [10]. Adding a gate to these devices to enable volatile or non-volatile voltage-controlled anisotropy will certainly be critical to enhance the computational capabilities of SOT-based architectures. To further enlarge the resistance change of MRAM and improve the scaling, researchers have put the effort into investigating novel crossbar array architecture (illustrated in Fig. 2) [7], as elaborated in Section 6.2.

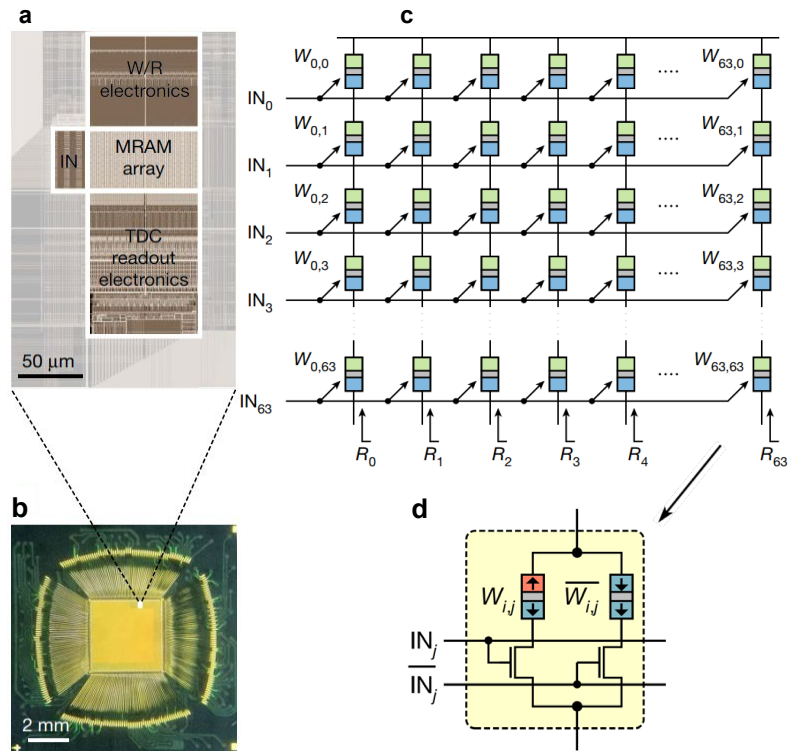

**Figure 2.** Layout (a) and micrograph (b) and of the 64 × 64 MRAM crossbar array. (c) MRAM crossbar array architecture and (d) configuration of each bit-cell. Reprinted with permission from [7]. Copyright 2022, Springer Nature.

**Concluding Remarks**

In conclusion, spintronics offers compelling opportunities for advancing neuromorphic computing by offering a range of bio-plausible hardware solutions. Various spintronic artificial synapses and neurons, driven by diverse physical mechanisms such as SOT, DWM, and magnetic skyrmions (as discussed in Section 1.4), have been effectively demonstrated. These advancements hold the potential for seamless integration into comprehensive brain-inspired spintronic systems. Nevertheless, several challenges remain to be addressed such as increasing the read-out signal, further investigation of bio-realistic neural devices, coupling control of neurons, and large scaling of compact and energy-efficient artificial neural networks. The state-of-the-art spintronic technologies have been discussed to meet these challenges including spintronic memristors, S-MTJ, STNO, spin-diodes, and new design of MTJ-based MRAM architecture. Spintronic neuromorphic computing is currently a technologically fast evolving field. The experimental demonstration of spintronics-based network-level neuromorphic computing remains to be further explored and to be implemented into large-scale hardware neural networks.

**Acknowledgements**
The work is supported partially by the SpOT-LITE Programme (A*STAR grant, A18A6b0057) through RIE2020 funds, the National University of Singapore Advanced Research and Technology Innovation Centre (A-0005947-19-00), National Research Foundation (NRF) Singapore (NRF-000214-00), Samsung Electronics' University R&D Programme, the project PRIN 2020LWPKH7 funded by the Italian Ministry of University and Research and by the European Union's Horizon 2020 research and innovation program under grant RadioSpin No 101017098 and under grant SWAN-on-chip No. 101070287 HORIZON-CL4-2021-DIGITAL-EMERGING-01.

## 1.3- Spin wave-based computing

Florin Ciubotaru, IMEC, Leuven 3001, Belgium (florin.ciubotaru@imec.be)
Andrii Chumak, Faculty of Physics, University of Vienna, Vienna 1090, Austria (andrii.chumak@univie.ac.at)
Azad J. Naeemi, School of Electrical & Computer Engineering, Georgia Institute of Technology, Atlanta, GA 30332, USA (an42@gatech.edu)
Sorin D. Cotofana, Quantum and Computer Engineering Department, Delft University of Technology, Delft, CD 2628, Netherlands (S.D.Cotofana@tudelft.nl)

**Status**

Magnonics [1] is an emerging solid-state physics field where Spin Waves (SW) – the collective excitations of the magnetic orders – and their quanta magnons are utilized, instead of electrons, for information transport and processing [2]. SW characteristics, e.g., GHz to THz frequency range, down to atomic scale wavelengths, pronounced non-linear and non-reciprocal phenomena, tunability, low-energy data transport and processing, offer many avenues towards building SW based nanoelectronics. The applied magnonics field is intensively growing, while SW sensing and SW radio frequency applications, which are becoming increasingly important, in view of the 5G technology requirements, are still in early stages of development. Boolean and unconventional SW computing have reached many milestones in recent years, and are experiencing constant growth [1,2]. Moreover, quantum magnonics attracts increasing attention within the community [2] and potentially offers an additional entanglement-related degree of freedom for quantum computing. The most important Boolean computing relevant achievements are the experimental realization of the inline majority gate [3], directional coupler, and magnetic half adder [4], as well as the basic circuits demonstration by means of micromagnetic simulations [5] (see Fig. 1 (a)-(c)). Magnon-based unconventional computing [2] is primarily associated with neuromorphic computing [6,7] although versatile approaches of wave-based computing including spectrum analysis or pattern recognition with magnonic holographic memory devices [8] can be placed in the same category (see Fig.1 (d)).

The concept of inverse-design magnonics, which given a certain functionality utilizes a feedback-based computational algorithm to obtain the corresponding device design [9], has been successfully utilized for radio-frequency applications [9] and neural networks [7]. Such an approach produces a device with rectangular ferromagnetic functional region patterned with square-shaped voids, as depicted in Fig. 1 (e). To demonstrate the universality of this approach, linear, nonlinear, and nonreciprocal magnonic functionalities were explored and a magnonic (de-)multiplexer (see Fig. 1 (e)), a nonlinear switch, and a circulator have been designed. Machine learning-based inverse design has significant potential for any kind of data processing, including the realization of complex multi-bit Boolean logic gates or neuromorphic networks [7].

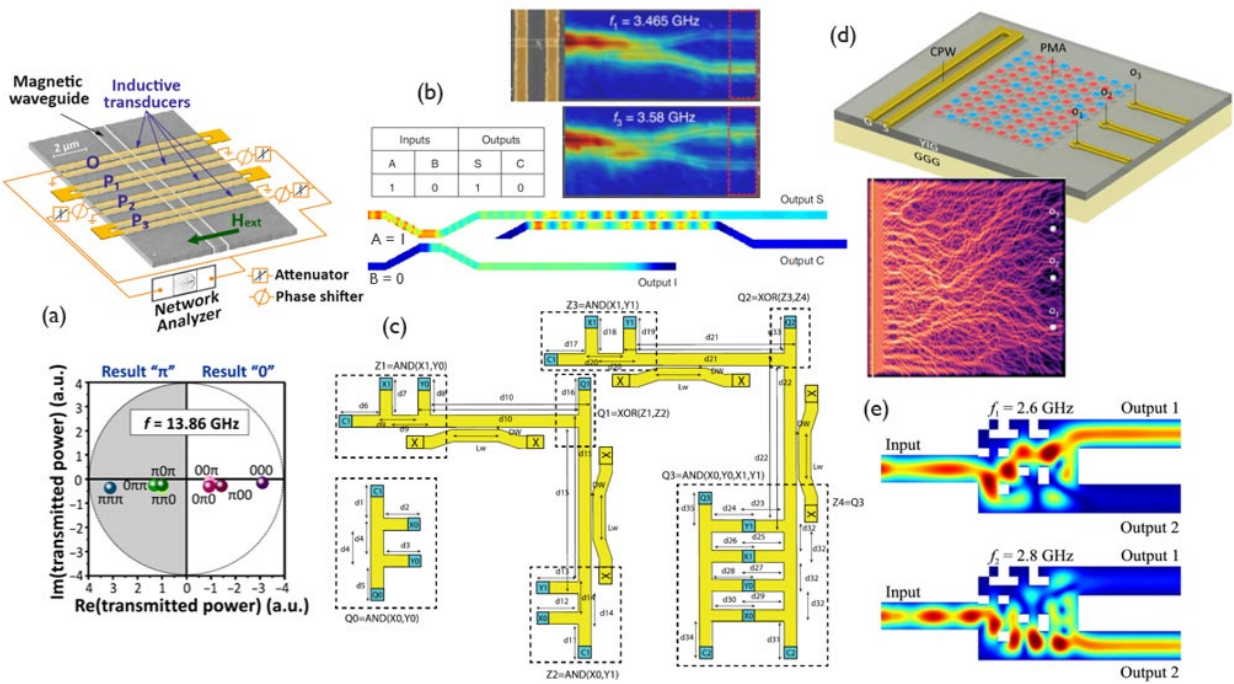

**Figure 1.** (a) Scanning electron micrograph of a sub-micron scaled spin wave majority gate (top) and the polar plot of the transmitted power for different input phases demonstrating strong and weak majority signals (bottom). Figure adapted from Ref. [2]. (b) 2D Brillouin light scattering spectroscopy maps of the spin-wave intensity recorded in a directional coupler for two excitation frequencies (f = 3.465 GHz and f = 3.58 GHz) (top). Operational principle of a magnonic half-adder demonstrated by micromagnetic simulations. Figure adapted from Ref. [4]. (c) 2-bit Inputs Spin Wave Multiplier. Reprint from Ref. [5]. (d) Schematic of a nanomagnet based spin-wave scatterer (top), and a spin-wave intensity pattern (bottom) for neural network applications. Reprint from Ref. [7]. (e) Magnonic demultiplexer device simulated by inverse design micromagnetic simulations. Reprint from Ref. [9].

## Current and Future Challenges

Various reported simulations and experiments have clearly demonstrated that SW could serve as information carriers and their interaction for data processing. However, the design of a fully magnetic computing system is far from being possible until effective solutions are found for: (i) constructing circuits out of magnonic gates and (ii) the realization of magnonic memories. Typical circuit construction challenges include fanout achievement and gate cascading, and they have been recently addressed in [5]. However, while the proposed approaches [5] enable magnonic circuit realization by enabling up to a fanout of 4 and direct gate cascading within the magnonic domain, they are expensive in terms of area and delay. Thus, to unleash magnonic computing full potential more effective solutions are required. Unfortunately, except the Holographic Memory concept introduced in [8], which is not a real magnonic memory as it doesn't store data in magnons, very little progress has been made towards the conceptual realization of magnonic memories.

Moreover, building magnonic computing chips requires many more components apart of simple magnetic waveguides used for SW propagation. Passing the input data from CMOS (the charge domain) to the magnonic circuitry requires high energy efficient scaled transducers. Typical SW transducers, e.g., inductive antennas, spin-transfer- or spin-orbit-torque based magnetic tunnel junctions, are scalable but very inefficient in terms of energy consumption. Their energy-delay characteristics need to be improved by few orders in magnitude to compete with the CMOS computing circuitry counterpart. Voltage driven transducers, e.g., based on magnetoelectric effects, might reach the required energy efficiency, yet such performances should be experimentally demonstrated at the nanoscale. Next to the transducers, the magnetic conduits should transport the information with minimum losses and delay. Recent studies demonstrated SW propagation lengths in scaled waveguides in a several micrometers range. However, not all the studied materials (e.g., $Y_3Fe_5O_{12}$) are CMOS compatible. Alloys based on CoFe(B) are widely utilized in MRAM technology and are promising candidates for magnonic conduits. Nevertheless, further material developments and optimizations, especially for the voltage driven transducers, will be required. In addition, SW amplifiers might be needed to restore the amplitude loss during the propagation if the circuit length exceeds the wave mean free path. These amplifiers also need to operate at very low energies (towards aJ), which suggest that the amplification process should also rely on voltage driven mechanisms, e.g., using Voltage Control of Magnetic Anisotropy (VCMA) or other magnetoelectric effects [10].

Magnonic circuit layout design is much more challenging than the one of a charge-based counterpart. SW propagation and interaction are quite sensitive to waveguide dimensions and geometries, e.g., SW behavior in straight waveguides and around corners are different due to reflections, and, as such, layout can significantly influence circuit performance and even make it malfunctions. Moreover, due to SW amplitude decay and dephasing phenomena the circuit size (chip real-estate) should be minimized by enabling 2D signal crossing and/or 3D interconnect.

**Advances in Science and Technology to Meet Challenges**

Much progress has been done in material development, realization and characterization of nano-scaled SW conduits below 100 nm, as well as in understanding the underlaying physical mechanism of SW generation and propagation, both in linear and non-linear regimes [2]. Most of the studies focus on SW properties (2D or 3D systems) and rely on optical characterization and/or micromagnetic simulations. However, the main challenge for building efficient magnonic computing circuits is related to the physical realization of an efficient energy coupling interface for the information transfer between electric and magnetic domains. The research progress on heterogeneous integration of multiferroic materials, magnetoelectric composite [10] and VCMA stacks for the generation and detection of SW could allow for the demonstration of nanoscale cascaded logic gates in a full electric experiment. Furthermore, the coupling of phonons or photons to SW could bring additional functionalities and enhance or control their characteristics, e.g., the group velocity or amplitude.

State of the art SW-based computing assumes phase encoding of information and is performed via not input-output format coherent majority gates. The direct cascading of such gates results in input data dependent circuit malfunctions. Despite of the fact that gate cascading solutions have been proposed they are rather expensive and induce large gate delay overhead, e.g., the gate delay is increased from few ns to more than 20 ns [5]. As such, to build competitive SW circuits more effective cascading schemes are required and potential solutions might be found by means of inverse-design [7,9] and/or by investigating alternative information encodings that may result in directly cascadable gates, and potentially enable the realization of more computation within one single gate. Given that information can be encoded in SW phase, amplitude, frequency, and any combination of those, a plethora of alternative and more effective SW computation paradigms could be potentially developed. Note that moving from charge-based circuits, e.g., CMOS, to SW circuits requires changes into the computer aided design framework. The traditional Boolean algebra-based logic synthesis should accommodate a new universal gate set composed of majority gate and inverter. Given that magnonic circuits are expected to be hybrid (to be interfaced with the environment within which they operate) and may include

CMOS parts, a not yet existing mixed micomagnetic-SPICE simulation framework should be developed. The SW circuit layout design is governed by completely different principles and has fundamental implications on the circuit performance and behavior. Thus, a novel approach to produce correct by construction SW circuit layouts is essential for the proliferation of the SW computing paradigm.

**Concluding Remarks**

Spin waves demonstrated to possess a high potential for computing but also for other emerging sensing or radio-frequency applications. The deep understanding of the associated physical phenomena and the development of materials and device fabrication techniques will allow the transition from the fundamental research to engineering devices in a near future. However, for computing applications, the SW devices should be integrated in hybrid magonic – CMOS architectures, where the magnonic units are utilized to solve some specific and energy consuming tasks. In this quest, several key components are still to be developed*,* as explained in this roadmap. Independent on the computing paradigm, an efficient information transfer interface between CMOS and magnonic circuits is the enabling factors towards real technologie. Development of multi-physics and SPICE design tools for device simulation as well as for circuit layouts will further pave the way for applications. Last, but not least, novel computing architectures as (spiking) neuronal networks based on interference, non-linear effects or non-reciprocity of spin waves could be developed for special applications.

**Acknowledgements**
The work of F. Ciubotaru and S.D. Cotofana was supported by the European Union's Horizon 2020 Research and Innovation Programme through the FET-OPEN project CHIRON under Grant Agreement 801055 as well as from its Horizon Europe research and innovation program within the project SPIDER (grant agreement no. 101070417). F. Ciubotaru acknowledges the support from imec's industrial affiliate program on Exploratory logic. A. Chumak acknowledge the financial support by the European Research Council (ERC) Proof of Concept Grant 101082020 5G-Spin and by the Austrian Science Fund (FWF) via grant no. I 4917-N (MagFunc).

## 1.4 - Computing with skyrmions

Riccardo Tomasello, Department of Electrical and Information Engineering, Politecnico di Bari, 70125 Bari, Italy (riccardo.tomasello@poliba.it)

Christos Panagopoulos, Division of Physics and Applied Physics, School of Physical and Mathematical Sciences, Nanyang Technological University, S637371, Singapore (christos@ntu.edu.sg)

Mario Carpentieri, Department of Electrical and Information Engineering, Politecnico di Bari, 70125 Bari, Italy (mario.carpentieri@poliba.it)

**Status**

Magnetic skyrmions are localized whirling spin-textures with topological properties and particle-like characteristics[1]. Research has exploded in the last decade with proposals for new materials and device concepts[1], offering intriguing functionalities ranging from memory to computing applications. Originally observed at low temperatures in B20 compounds, today, skyrmions can be found in many thin film materials (ferro-, ferri-, and synthetic antiferro-magnets (SAFs)) at room temperature[1]. Their small size and the possibility to easily manipulate them electrically make these topologically protected chiral spin configurations attractive as information carriers in compact, and energetically efficient devices. Furthermore, the anticipated insensitivity to defects and potential for low energy consumption[2] have accelerated efforts to understand their formation, stability, and dynamics sufficiently well, already extending the interest towards unconventional applications. In particular, the topological properties of magnetic skyrmions offer new paradigms in reservoir, stochastic, neuromorphic, and quantum operations[3], [4].

Single skyrmions or a skyrmion fabric [5] are promising for reservoir computing because the non-linear dynamics of magnetic skyrmions can increase the systems nonlinearity and therefore the efficiency of the reservoir. In stochastic computing, the essence for optimal operation is the decorrelation of the bitstreams[6]. This feature can be achieved by a skyrmion-based reshuffle chamber (a missing element in current implementations of stochastic computing) that has been demonstrated taking advantage of the diffusion properties of skyrmions [6] (Fig. 1a).

Skyrmions have been also proposed for neuromorphic application. While skyrmion-based neurons are still only theorized[3], a skyrmion-based synapsis has already been experimentally demonstrated[7] (Fig. 1b), with the weight represented by the Hall resistivity.

Skyrmions could also serve as a source for non-Abelian statistics. Imprinting skyrmions on superconductors may trigger the formation of special Majorana quasiparticles, granting unrivalled resilience to the decoherence problem that plagues other quantum computing platforms[8]. Nano-skyrmions are also of interest for their potential as a logical element of a quantum processor[8]. They develop quantized eigenstates with distinct helicities and out-of-plane magnetizations. In a skyrmion qubit, information is stored in the quantum degree of helicity, and the logical states can be adjusted by electric and magnetic fields, offering an operation rich regime with high anharmonicity.

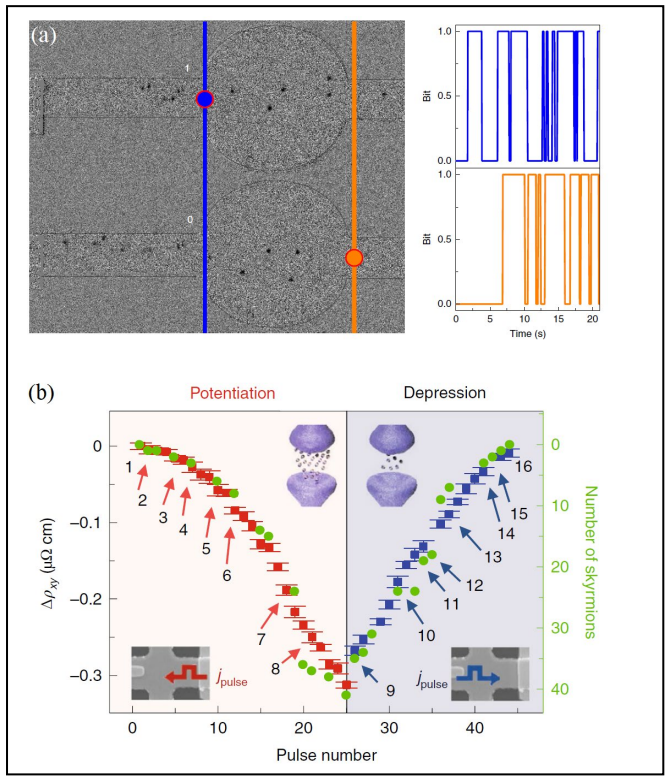

**Figure 1.** (a) example of a skyrmion reshuffler. Reproduced with permission from Ref. [6]. (b) memristive behavior of skyrmions for synapsis applications. Reproduced with permission from Ref. [7].

**Current and Future Challenges**

Fundamental challenges for utilizing magnetic skyrmions in technology include controlling their size, sustaining positional stability, enhancing electrical readout, deterministic nucleation and annihilation, reducing/suppressing the skyrmion Hall angle while maintaining high velocity, but also achieving an overall integration with digital circuits and associated circuit overhead.

Skyrmion position can be controlled by engineering pinning sites with lithography or ion irradiation[1]. Improvements in electrical readout calls for the use of optimized large-Tunneling Magnetoresistance Magnetic Tunnel Junctions (TMR-MTJs)[9]. Nucleation/annihilation protocols enable deterministic functionalities[1], but should be optimized to decrease energy consumption. The skyrmion Hall angle can be suppressed in SAFs[2]. However, the skyrmion velocity[10] is still far from theoretical predictions, calling for new and/or optimized materials and device architectures.

On the side of unconventional applications, Reservoir computing based on magnetic textures has not been realized experimentally yet. In stochastic computing, the first proof of concept of an algebraic computation based on skyrmions is still missing and hence a full skyrmion-based implementation. Furthermore, energy efficiency, velocity and accuracy in computation should be evaluated and compared with standard CMOS systems. In neuromorphic computing, some proposals rely on the use of a skyrmion-based spin-transfer-torque (STT) oscillator which has yet to be demonstrated. Whereas proposals involving memristive skyrmion behavior, such as skyrmion synapsis, should be based on large TMR-MTJs for improved detection and compact solutions. Specifically for the results in [7], the accuracy of the network could be improved, compared to the current estimate of 89%.

Beyond non-interventional creation or observation studies, skyrmions promise to dramatically improve quantum operations. Skyrmion-vortex interaction in device architectures of imprinted magnetic skyrmions on conventional superconductors can assist topological quantum computing by operations carried out on Anyons. A novel quantum hybrid architecture composed of Néel skyrmions and Niobium grants realistic hope[11]. Moving forward, in a homogeneous chiral magnet and superconductor stack, a skyrmion-vortex pair – and hence a Majorana zero mode – would be intrinsically mobile. This allows for non-perturbative, non-contact braiding operations by moving skyrmion-vortex pairs around the surface.

Quantum skyrmions in frustrated magnets offer a new element for the construction of qubits based on the energy-level quantization of the helicity degree of freedom. The skyrmion state, energy-level spectra, transition frequency, and qubit lifetime are configurable and can be engineered by adjusting external electric and magnetic fields, offering a rich operation regime with high anharmonicity.

**Advances in Science and Technology to Meet Challenges**

Over the next two decades, a concerted experimental effort will promote skyrmions to future devices, hence, realizing their technological potential for information processing transcending existing limits. Challenges, such as decreasing energy consumption of skyrmion devices, controlling the skyrmion size, suppressing the skyrmion Hall angle, could be achieved by exploring new materials architectures. We can reduce dimensionality with emphasis on 2D van der Waals materials. At the same time, we can explore 3D bulk materials that, thanks to improved imaging techniques, demonstrate the presence of static 3D structures, such as vortex rings and hopfions[12]. Whereas, on the dynamical side, so far the community has been relying on theoretical predictions of the current-driven Hopfion motion.

Emphasis on frustrated magnetism, intrinsic to triangular or hexagonal lattices with antiferromagnetic spin correlations, can utilize the induced non-collinear magnetic order, which itself breaks spatial inversion symmetry for the formation and manipulation of skyrmions only a few nanometers wide. We can also focus on the design of hybrid systems by combining ferro-, ferri-, and antiferro-magnets. Specifically for unconventional skyrmion applications, experimental realization of reservoir computing needs efficient and precise electrical measurements among multiple contacts for a reliable resistance evaluation linked to the magnetic texture distribution. Stochastic computing demands an integration with CMOS technology via stacks compatible with STT-MRAM technology already integrated and commercialized. Neuromorphic computing calls for the enhancement of skyrmion detection (Fig. 2a) beyond the 30% TMR threshold for MTJs. This could be achieved by combining state-of-the-art CoFeB-MgO MTJ with conventional skyrmion-hosting magnetic multilayers. Another direction could be combinatorial, including topological magnetism and acoustic waves, already promising for skyrmion-based synaptic behavior [13].

For quantum computing operations in skyrmion-superconductor hybrids, major tasks need to be performed for device capability. Firstly, ensuring that the magnetic interactions can spin-polarize the superconductor and cause a topological phase transition. Secondly, for Majorana braiding, the magnetic homogeneity of the architecture needs to be better than commonly achieved using magnetron-sputtering. In qubit-technology hardware, the applicability of nano-skyrmions can be further improved with the development of cleaner magnetic samples and interfaces in engineered architectures, without trading off qubit anharmonicity and scalability. Notably, demonstrating tunable macroscopic quantum tunneling, coherence, and oscillations for magnetic nano-skyrmions will also establish helicity in topologically protected chiral spin configurations as a quantum variable; much like macroscopic quantum tunneling and energy quantization in Josephson junctions, thus the fundamental physical step for developing a practical skyrmion qubit.

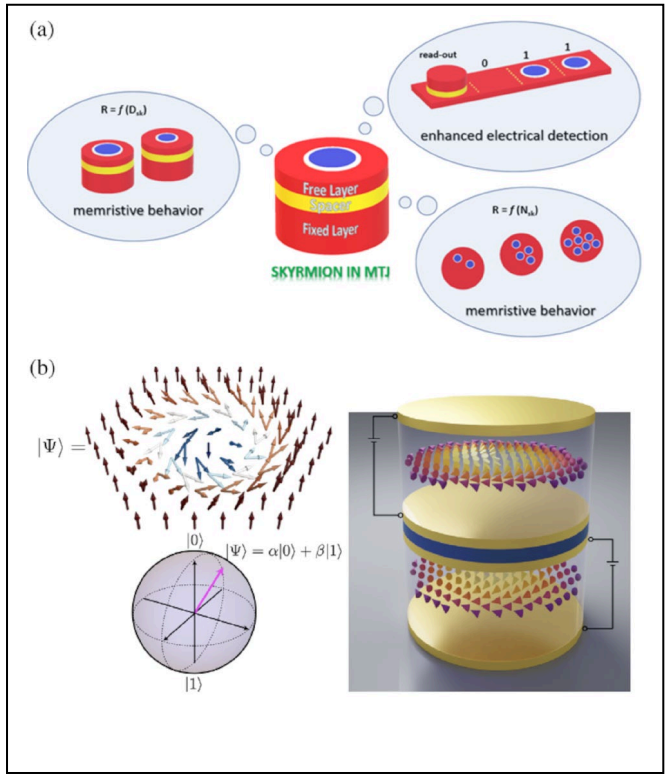

**Figure 2.** (a) Sketch of the improvements of skyrmion detectivity through the use of a skyrmion-based MTJ. R is the electrical resistance of the MTJ, $D_{sk}$ is the skyrmion diameter, and $N_{sk}$ is the number of skyrmions. (b) Skyrmion qubit concept. Reproduced with permission from Ref. [8].

**Concluding Remarks**

Skyrmions are fascinating topological magnetization configurations with realistic potential for a beneficial impact on computing paradigms. Their small size, particle-like behavior, topological properties, manipulability by electrical current, as well as memristive features promise disruptive advances in reservoir, stochastic, neuromorphic, and quantum computing. Stabilizing skyrmions in large TMR-MTJ will enhance the skyrmion detectivity with unprecedented effects on unconventional operations. Extending the investigation to 3D bulk magnetic systems and 2D materials will lead to major breakthroughs and new functionalities associated with complex topology and multiple degrees of freedom. Remarkable advancements in unconventional computing can be driven by the combination of different physical systems, such as topological magnetism and acoustic waves which already promise to control skyrmion-based synaptic behavior. Skyrmions interacting with superconductors can lead to chiral superconductivity and Majorana braiding platforms. Whereas, nano-skyrmions stabilized in magnetic disks bounded by electrical contacts will allow static fields to control the quantized energy spectra, enabling changes in the helicity between energetically favored levels. It is expected that skyrmions will be a major building block for the next generation of low power computing architectures, transcending from the classical to the quantum regime.

**Acknowledgements**

This work was supported by the Project No. PRIN 2020LWPKH7 funded by the Italian Ministry of University and Research. C. Panagopoulos acknowledges support from the National Research Foundation (NRF) Singapore Competitive Research Programme NRF-CRP21-2018-0001, and the Singapore Ministry of Education (MOE) Academic Research Fund Tier 3 Grant MOE2018-T3-1-002.

## 2.1 – Neuromorphic computing with memristive devices

Peng Lin, College of Computer Science and Technology, Zhejiang University, Hangzhou, 310013, China (penglin@zju.edu.cn)
Gang Pan, College of Computer Science and Technology, Zhejiang University, Hangzhou, 310013, China (gpan@zju.edu.cn)
J. Joshua Yang, Department of Electrical and Computer Engineering, University of Southern California, Los Angeles, CA 90089, USA (jjoshuay@usc.edu)

**Status**

Neuromorphic computing is a promising paradigm of artificial intelligence (AI) systems that aims at developing an efficient computing architecture with great physical and functional resemblance to the biological brains. Memristors, which represent a group of emerging memory devices, can reversibly change their conductance under a variety of physical switching sources such as phase change, ionic motion, ferroelectricity and ferromagnetic switching [1]. In particular, the ionic memristors shares some similarities at the physics level with ion transport in nerve cells (e.g., $Ca^{2+}$, $Mg^{2+}$, $Na^+$, $K^+$), which equips them with desirable ion dynamics, enabling a variety of more efficient emulations of neuronal and synaptic functions for neuromorphic computing [2].

In recent years, memristors with decent array sizes have been used as static synapses for physical implementations of artificial neural networks (ANNs), which are low hanging fruits for memristors because neural network applications take advantages of the strengths and avoid the weaknesses of typical memristive devices as revealed in Ref [3]. This trend is reflected by a significant increase in application-oriented publications since 2018 (fig. 1). Each memristor is not only a weight storage unit, but can also directly process weighting function to upstream voltage inputs in the form of voltage-conductance multiplications [4], and thus co-locates the memory and processing functions within the same cell. A memristor array is essentially a physical neural network in between two neuron layers with many possible array arrangements, and can be extended to 3D layout to implement complex neural networks [5]. Owing to its low power, highly parallel computing paradigm, memristor based systems have found numerous successes in ANN related applications and demonstrated excellent computing efficiency exceeding 100 TOPS/W.

In the meantime, dynamical neuronal and synaptic properties of memristors are also under extensively study to provide more capable and compact building blocks for neuromorphic computing. In ANN applications, linear conductance modulation of memristors using a burst of identical pulses have been reported [1], which shows promises to implement accurate weight updates for fast on-chip training of ANNs. Meanwhile, functional memristive devices were developed to implement spiking neural network (SNN) – a potentially more efficient neural network with a high bio-plausibility. Important neuron models such as leaky-integrate-and-fire (LIF), Hodgkin-Huxley (HH) and plastic synaptic behaviors such as paired-pulse-facilitation (PPF) and spike-timing-dependent-plasticity (STDP) have been achieved by harvesting more dynamical behaviors of the memristors. These novel functions were achieved in a very compact form normally with a couple of memristive devices instead of bulky circuits with many transistors in pure CMOS implementations. However, the overall scales of these demonstrations were still limited to small arrays, which is significantly hampering the overall capability of neuromorphic computing systems in practical applications.

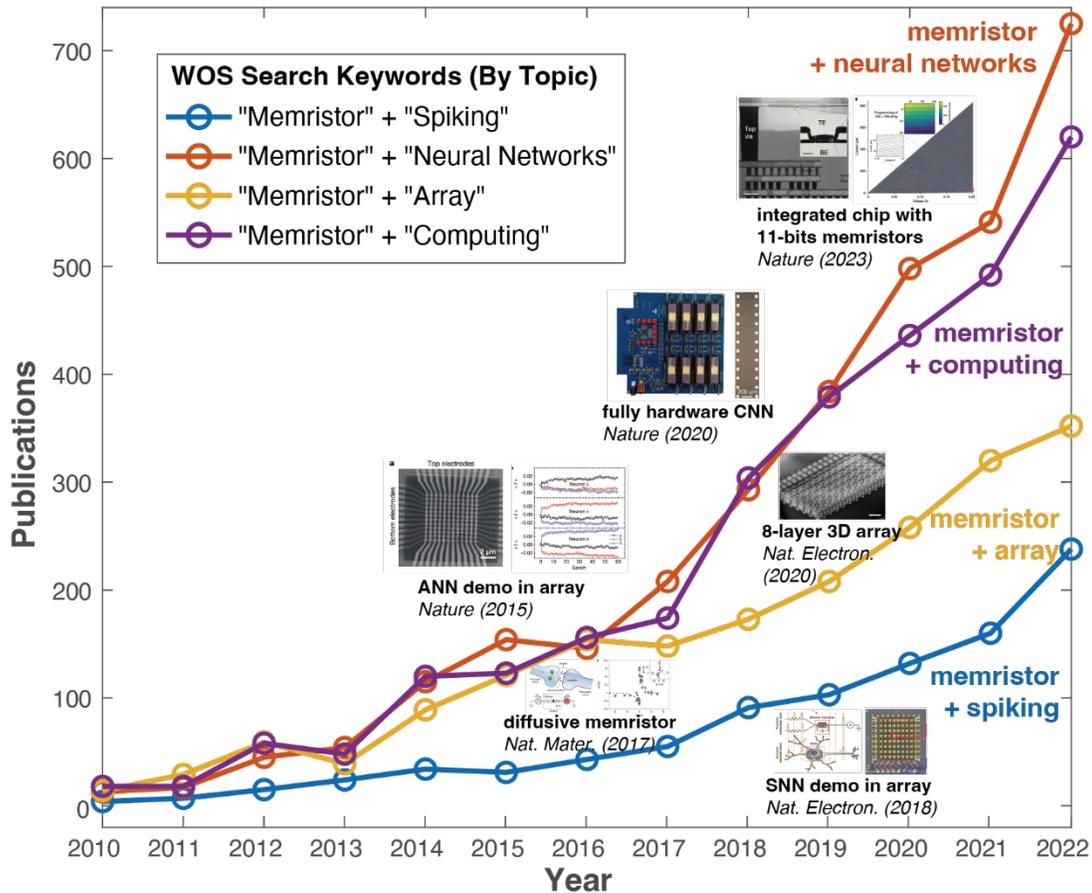

**Figure 1. Number of publications related to memristor based neuromorphic computing system, retrieved from Web of Science (WOS) database, data source: Clarivate.** Inset: ANN demo in array [4]; diffusive memristor [2]; fully hardware CNN [6]; SNN demo in array [7]; 8-layer 3D computing array [5]; integrated chip with 11-bits memristors [8].

**Current and Future Challenges**

Even though building a neuromorphic system requires a synergistic effort from both hardware and software, the device performance of memristors still plays a decisive role in determining the ultimate capability and functionality of the neuromorphic system. Currently, there are a few prominent challenges for memristive devices.

First, it is still challenging to build a large-scale memristor arrays without an access transistor. At present, the majority of the memristor chips are based on the so-called one-transistor-one-memristor (1T1R) cell design, for which a transistor is integrated with a memristor and served as a current regulator for the memristor cell. The access transistor can (1) suppress the leakage current from the unselected cell, (2) use current compliance to achieve accurate conductance tuning and mitigate switching variations. However, the downside of having an access transistor is that it essentially limits the use of any array-wise parallel programming strategies, and raises challenges in designing an asynchronized system, such as those based on SNNs. The 2D scalability and 3D stack-ability of the memristor arrays are also limited by the addition of a transistor in each cell.

Secondly, linear analog conductance modulation using identical pulses is a key requirement for on-chip training of ANNs. Although it has been demonstrated in small prototype arrays, uniform linear conductance modulation across large-scale arrays is still challenging to achieve because device-to-device variations of switching voltages, conductance range, response time, etc. can all affect the conductance modulation process. In addition, it is also highly desirable to have symmetric programming for potentiation and depression, which, however, is not well supported by most of the memristive switching mechanisms.

Lastly, the variability issue in memristors is also a key limiting factor for the implementation of SNNs since dynamical functions such as STDP and LIF usually have lower tolerance for variations. For example, a standard STDP response of synapses is nonlinearly related to the timing differences between the pre- and post-synaptic spikes. As a result, variations in the time domain (such as inaccurate firing delay from the LIF neuron) would be nonlinearly magnified in synaptic response, causing large errors during learning. Finding a solution to improve the scalability of these SNN functions, whether through more precise control of ionic motions, or through other compensation strategies, are highly desired for the development of SNN based neuromorphic system.

Of course, challenges are also existed in other aspects of a neuromorphic systems. Developing a global training algorithm for large scale SNNs is among the top of these challenges. Moreover, implementing neuromorphic systems still requires more efficient peripheral circuit designs, more sophisticated system architecture, and would require a better understanding of the working principles of SNNs.

**Advances in Science and Technology to Meet Challenges**
First, we would like to see continuous efforts in material and device engineering develop new device concepts with breakthroughs in device performance and gain better understanding of the switching mechanisms. This idea is supported by rich switching behaviors from different types of memristors, which are dictated by a combination of factors including material compositions, fabrication processes, device morphologies and many others, therefore provide a high degree of design freedom to tailor a device under specific application requirements. Alternative memristor design based on non-filamentary switching mechanisms may be developed to achieve improved switching uniformity. For example, a new type of electrochemical memristor was reported, and demonstrated better switching uniformity and excellent analog programmability, though its application in large arrays needs further verification [9].

Meanwhile, we hope to see new fabrication technologies or material synthesis methods for memristors, such as using sophisticated tools from commercial foundries. It is known that the use of ion implantation for CMOS process provided dramatic improvement to the doping profile of MOSFET. We believe that a disruptive improvement may also be achieved in a similar effort. For example, in a preliminary study, epitaxial tool was used to grow single crystal SiGe film with nanometer wide dislocation channels, which acted as predefined ion channel for switching. Owing to better control over the ion transport, improvement in switching uniformity of memristors was achieved [10].

Finally, challenges at device and hardware level may also be overcome through complementary research effort in computer science and neuroscience domains. For example, more robust, hardware-friendly algorithms and computational models could be developed to mitigate the variability issues of memristors and utilize some unexpected properties discovered in new device exploration. Meanwhile, co-optimizations of the parameters for both memristive devices and neural networks could be achieved through comprehensive modeling and simulations. Lastly, our understanding of the brain is still at its infant state, it is also possible that new findings in neuroscience could help to establish new training methods and design new network architectures.

**Concluding Remarks**
Neuromorphic computing is a disruptive technological solution to future AI, and memristor is one of the leading candidates to implement parallel, analog and in-memory computing as well as rich dynamics inside neuromorphic computing systems. At the current stage, large-scale memristor based neuromorphic systems are mainly based on ANN algorithms, while SNN based demonstrations are far behind, primarily due to lack of appropriate training algorithms and the challenges to reliably obtain the desirable dynamical functions at large scale.

As more technical challenges described in this roadmap being resolved, it is expected to see a much more substantial progress made in neuromorphic computing. A large-scale memristor system based on a comprehensive SNN design can lead to significant improvements in energy efficiency, performance, and functionality over existing AI hardware. Meanwhile, the SNN hardware could, in return, inspire the development of SNN algorithms or even the understanding of biological neural networks, which have been inefficient to simulate using conventional computers.

**Acknowledgements**

This work was partially supported by the National Science Foundation under contract No. 2023752, Natural Science Foundation of China (No. 61925603), The Key Research and Development Program of Zhejiang Province in China (2020C03004) and Major Program of Natural Science Foundation of Zhejiang Province in China (LDQ23F040001).

## 3.1– Nanomaterials for unconventional computing

Aida Todri-Sanial, Electrical Engineering Department, Eindhoven University of Technology, The Netherlands (a.todri.sanial@tue.nl)

Gabriele Boschetto, Microelectronics Department, LIRMM, Université de Montpellier, 34095 Montpellier, France (gabriele.boschetto@lirmm.fr)

Kremena Makasheva, Laplace, Laboratory on Plasma and Conversion of Energy, CNRS, UT3, INPT, University of Toulouse, 31062 Toulouse, France (kremena.makasheva@laplace.univ-tlse.fr)

**Status**

With the emergence of nonconventional computing paradigms, one has the means to overcome the fundamental limitations of von Neumann architecture and perform highly complex functions with extremely low power. This prompts for materials and devices that can emulate the biological functions of neurons and synapses. Combining memory and resistor, memristors have become the most important electronic component for brain- inspired neuromorphic computing. The device has the ability to control resistance with multiple states by memorizing the history of previous electrical inputs – allowing it to mimic biological synapses and neurons of the biological neural networks. The switching in memristor devices is a reversible and controllable change of resistance induced by different stimuli, such as current, voltage, or light, with different physical processes such as ionic/electronic motion and redox reactions. Thus, the material selection plays a key role in the conductive path formation and modulation of the resistive switching behavior, and here we review them based on material properties. Owing to the dependence of their resistance states in the history of the applied electrical bias, memristors can store information in the form of electrical resistance and are typically driven by one of the four main mechanisms: electrochemical reactions (namely, redox and ion migration), phase changes (such as thermally activated amorphous-crystalline transitions), tunnel magneto-resistance (as such as spin-dependent tunnel resistance) or ferroelectricity (namely, tunneling or domain-wall transport). In addition, memristors can allow to process information intrinsically through the "let physics compute" (namely, perform complex signal transformation with physical dynamics), which are beneficial beyond neuromorphic computing, such as solving NP-hard optimization problems and hardware security. To uncover their potentiality, we summarize here the nanomaterials used for unconventional computing and their different types of switching mechanisms (Figure 1).

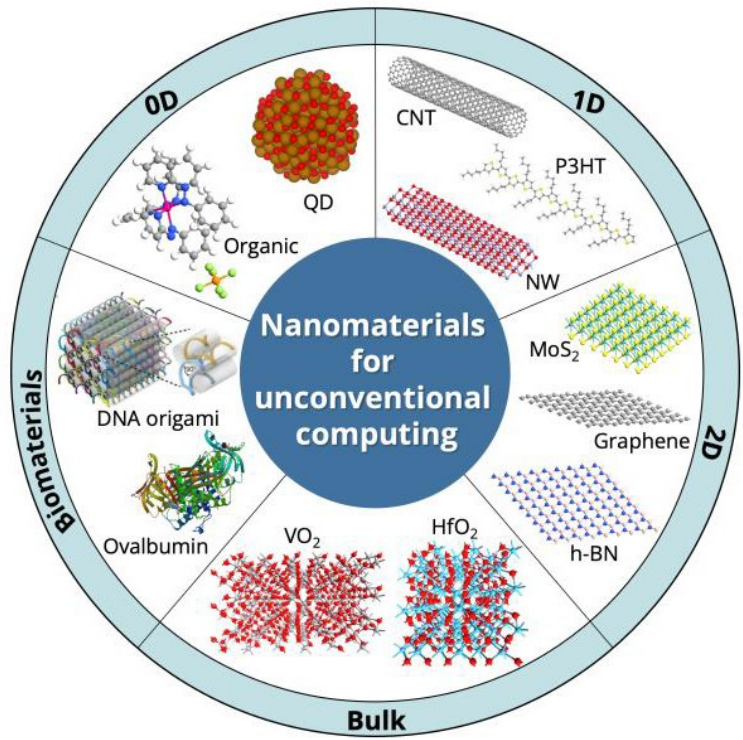

**Figure 1.** Illustration of different classes of nanomaterials used for unconventional computing, based on their dimensionality.

*0D Nanomaterials* – a variety of 0D materials have been investigated for memristors, mainly metal nanoparticles (NPs) and semiconductor quantum dots (QDs). Metal NPs are typically used to modify the resistive layer or electrodes to lower the charge injection barrier and interfacial potential drop of the electrode. Semiconductor QDs have also been investigated as promising candidates for developing electronic synaptic devices due to their electrical and optical properties. For example, charge accumulation in QD floating gate has obtained linearly programmable conductance states by controlling applied gate voltages. The optical properties of QDs enable electro-photoactive synaptic devices, offering a promising candidate for new electronic synaptic devices. In addition, pairs of QDs are explored to serve as a single basic element in a quantum logic device such as quantum bit or qubit. At the same time, tunable small organic molecules (e.g., azo-aromatics) and devices coupling ionic with electronic currents are investigated for emulating biological synapses behavior such as plasticity for continual learning. Resistance switching in most of these devices relies on either electrochemical doping, ion migration, or charge trapping mechanisms.

*1D Nanomaterials* – Carbon nanotubes (CNT) are among the most widely studied 1D nanomaterial with metallic or semiconducting behavior depending on their chiral vector. CNT network-based transistors have been demonstrated as a synaptic transistor with physical mechanisms, including charge trapping in oxide dielectrics, ion migration, and electric double layer effects in polymer dielectrics. CNT networks have been used to show simple realizations of plasticity learning for spiking neural networks. In addition, CNT-based transistors have been pursued for post-silicon digital logic and memory implementation. Semiconductor nanowires (NW) possess attractive attributes for neuromorphic devices. A wide range of inorganic 1D nanomaterials (metal oxides, Ag, and Cu nanowires) have shown both volatile and non-volatile resistive switching. In addition, organic polymer nanowires (e.g., P3HT) can emulate synapses' morphology and possess a learning mechanism similar to biological ion channels.

*2D Nanomaterials* – 2D materials, including graphene, transition metal dichalcogenides (TMDs), and hexagonal boron nitride (h-BN), have been widely investigated as emerging materials for low power transistors, sensors, and memristive devices. 2D material-based memristors can be further categorized based on their device geometry as lateral, vertical, and heterojunction structures. Depending on their geometry and materials, the switching mechanisms of these devices are based on phase transition, filament formation, charge trapping, defect migration, vacancy migration, or direct tunneling. Such devices are promising candidates for energy efficiency as artificial synapses while emulating plasticity for short-term and long-term memory.

*Bulk Materials* – Metal oxides exhibit electrical, optical, and semiconducting properties suitable for memristive devices. Interestingly, the crystal structure of some of these metal oxides undergoes changes under external stimuli such as thermal field, strain energy, surface energy, external force, magnetic field, and applied electric bias. This makes such materials suitable for memristor devices. For instance, resistive switching behavior from the repeatable formation of conductive filaments (i.e., oxygen-deficient or oxygen vacancy rich) with electric bias yields the low and high resistance states in oxide memristors such as hafnia ($HfO_2$). In addition, vanadium dioxide ($VO_2$), which shows a reversible metal-insulator transition (MIT) at near-room temperature, and a reversible large conductance change, has been used to emulate biological neurons (i.e., oscillating behavior) and synapses. $VO_2$ has also been used in field-effect transistors and gas sensors.

*Biomaterials* – These materials have attracted attention due to their long time natural evolution and well-defined "structure-function" relation. Proteins, in particular egg albumen and ferritin, have been shown suitable for developing memristors, with high plasticity in synaptic networks. In addition, the self-assembly of nucleic acids (DNA and RNA) has provided a powerful and effective approach for constructing synthetic molecular structures, tiles, 2D lattices, 3D crystals, finite 2D shapes, DNA origami, and more complex 3D nanostructures. Such DNA structures enable the engineering of molecular structures with programmable shapes and properties for applications such as drug delivery and biological computing.

**Current and Future Challenges**
In the Table below, we provide the current advancements in nanomaterials and their switching mechanisms, which lead to unique electrical, optical, magnetic, or quantum properties that are the basis for nonconventional computing paradigms.

| Material Dimensionality | Material Synthetic Route | Switching Mechanism | Materials (a few examples) | Switching Energy (fJ) | $I_{ON}/I_{OFF}$ | Reference |
|---|---|---|---|---|---|---|
| **0D** | Wet chemistry | Charge-trapping | Organic, *mer*-$[Ru(L)_3](PF_6)_2$ | 10pJ | - | [1] [2] |
| | Molecular beam epitaxy, ion implantation, wet chemistry, vapor-phase | Hybrid charge-trapping filament formation, multi-band emission | BPQD; QD | - | $10^1 - 10^7$ | [3] [4] |
| **1D** | Wet chemistry | Electrochemical doping | P3HT / PEO | 10 fJ | - | [1] |
| | CVD, hydro-thermal growth | Ion migration | ZnO NW | - | <6 | [5] |
| | CVD, arc discharge | Charge trapping | CNTFET with DMC | - | $5 \times 10^4$ | [6] |
| **2D** | CVD, PVD, exfoliation | Defect migration | Au-$MoS_2$-Au | - | $<10^4$ | [7] |
| | | Phase transition | Au-$MoS_2$-Au/Ti-$TaS_2$-Ti | - | $10^1 - 10^7$ | [7] |
| | | Vacancy migration | Pd/$WSe_2$/Pt; Pd/$WS_2$/Pt | >30fJ | $2 - 10^{10}$ | [7] |
| | | Filament formation | Cu/$MoS_2$/Au; Metal/*h*-BN/Metal | - | $2 - 10^4$ | [7] |
| | | Schottky/ Direct tunneling | Au/$MoS_2$/Au | - | $10^8$ | [7] |
| **3D** | Sputtering, CVD, hydrothermal growth | Electrochemical Redox | Metal oxides ($TiO_2$, $HfO_2$, $TaO_x$) | >10fJ | 2 - 40 | [8] |
| | | Phase change | Metal-insulator transition ($VO_2$) | >100fJ | <$10^3$ | [8] |
| | | Magnetic tunneling | Magnetoresistive materials (MgO) | >10fJ | 2 - 3 | [8] |
| | | Ferroelectric polarisation | Ferroelectric materials ($BiFeO_3$, $BaTiO_3$, $PbZrTiO_x$) | >100fJ | 45-300 | [8] |
| **Biomaterials** | Natural way and Wet chemistry for synthetic growth | Generation of conductive filament | Proteins (silk fibroin, ferritin, collagen, egg albumen) | - | 3 - $10^7$ | [9] |
| | | Biochemical operations; Self-assembly of nucleic acids | 2D and 3D DNA nanostructures | - | - | [10] |

**Advances in Science and Technology to Meet Challenges**

*Material and device variability and stability* – these characteristics rely on the intrinsic quality of the active materials and manufacturing process, which are yet to be mastered. This leads to some variability in the material properties due to, for instance, defect density and grain size. A major challenge is to reduce variability and enhance the reliability of material growth with high crystallinity and uniform thickness and effective passivation techniques without deteriorating device performances. The device characteristics can be precisely controlled to realize functional circuits based on the switching mechanism and material properties.

*Plasticity* – as in biological neural networks, communication between neurons is dynamic and occurs at different time scales. Communication strength depends on the history of synapse activity, also known as plasticity. Short-term plasticity facilities computation, while long-term plasticity is attributed to learning and memory. To enable on-chip learning, it is important that artificial synaptic devices display long-term memory and architectures to allow for local learning rules.

*Large scale integration* – to provide commercially available unconventional computing paradigms using nanomaterials (2D, QD, organic, etc.), it is important to achieve large-scale device array integrated with other circuits to show entire system operation while being CMOS compatible process.

*Energy consumption* – reducing the energy consumption of devices and electronics is an important endeavor for future low-power computing. It is not only important to develop low power devices but also the architecture of the full system in which it is implemented to achieve an energy efficient computing system.

*Biofriendly materials* - The use of biocompatible and biodegradable materials in electronic devices can be an important trend in the development of green electronics. Compared to metal-oxide semiconductors, biofriendly polymers and/or natural materials are attracting interest for their suitability on flexible neuromorphic platforms.

*Collaborations and training* – a close cooperation between neuroscientists, device physicists, computer scientists, computer architecture, and material scientists is of utmost importance to design and fabricate integrated circuits based on these new devices and realize the full potential of novel computing paradigms. The rapidly growing knowledge in each of these domains provides new insights and concepts to design energy efficient computing, necessitating collaborative efforts to train the new generation of students and scientists.

**Concluding Remarks**

Unconventional computing paradigms have propelled the research into novel devices that have led to a wide variety of solutions in terms of device physics, materials, and information processing. Despite the recent successes and advancements in novel devices and materials, more research is necessary to overcome the current limitations of devices to lower their variabilities and increase long-term operations, state retention, and modulation for enabling both short-term and long-term plasticity.

**Acknowledgements**
ATS and GB acknowledge the support from the EU H2020 SmartVista project with grant agreement No. 825114, EU H2020 NeurONN and Horizon EU PHASTRAC project with grant agreement No. 871501 and No. 101092096. KM acknowledges the support from the Agence Nationale de la Recherche in France, project ANR BENDIS (ANR-21-CE09-0008).

## 3.2 – Neuromorphic Computing with Emerging Two-Dimensional Nanomaterials

Vinod K. Sangwan, Department of Materials Science and Engineering, Northwestern University, Evanston, IL 60208, USA (vinod.sangwan@northwestern.edu)
Amit Ranjan Trivedi, Department of Electrical and Computer Engineering, University of Illinois at Chicago, Chicago, IL 60607, USA (amitrt@uic.edu)
Mark C. Hersam, Department of Materials Science and Engineering, Northwestern University, Evanston, IL 60208, USA (m-hersam@northwestern.edu)

**Status**

The emergence of two-dimensional (2D) and van der Waals (vdW) materials has invigorated fundamental research at the device level for brain-inspired computing hardware.[1, 2] A critical component of neuromorphic circuits is an analog non-volatile memory (NVM) that is not only fast, reliable, and high-density but also possesses multiple states and internal temporal dynamics to mimic the spike-based learning rules of biological synapses. Crossbars of NVM technologies based on conventional bulk materials, such as memristors, phase change memories, and magnetic and ferroelectric tunnel junctions, can outcompete CMOS counterparts for neural network performance metrics. All of these NVMs have also been realized using 2D materials with unprecedented functionalities (e.g., gate tunability) that translate into improved performance as a result of simplified circuit architectures. For example, 2D materials have been integrated into atomically thin vertical memristors with femtojoule switching energies (Fig. 1a,b).[3] The most promising vertical memristors are based on 2D transition metal dichalcogenides (TMDCs) or hexagonal boron nitride (hBN) where resistive switching has been achieved with intrinsic defects or metal cations. Although the constituent 2D materials can be grown over a wafer scale, most of the demonstrations thus far have been limited to 10 x 10 crossbar arrays without a selector (Fig. 1c,d).[3] A particularly promising approach is a self-selective crossbar based on two hBN memristors with volatile and non-volatile switching in an Au/hBN/graphene/hBN/Ag stack (Fig. 1c).[4] Although some applications have been proposed for 2D vertical memristors (e.g., RF switches, encryption circuits), their characteristics and functions are similar to conventional two-terminal memristive systems.[3]

To gain more unique functionality, semiconducting 2D materials (e.g., MoS2) can also be integrated into lateral memtransistors where nonvolatile switching is tuned by a third gate electrode (Fig. 1d,e).[5] In addition, 2D channels enable dual-gated control where one of the gates can achieve tunable learning behavior, while the other gate can be used as a selector in a manner analogous to a one-transistor-one-memristor (1T1M) crossbar (Fig. 1f,g).[6] Lateral memtransistors are also compatible with multiple electrodes to realize heterosynaptic learning behavior.[5] Memtransistors have been generalized to a wider class of heterojunctions using charge trap, floating gate, ferroelectric, conducting bridge, and phase change memories.[2] Crossbars consisting of 10 x 10 memtransistors have been experimentally demonstrated, achieving the same level of complexity as 2D vertical memristors (Fig. 1d,g).

Solution-processed 2D and vdW materials are also promising for printed and flexible neuromorphic circuits. For instance, femtojoule vertical memristors and memcapacitors have been demonstrated using printed films on flexible substrates.[7] However, most of these devices use electrochemical filaments such as Ag and Cu, and thus the role of the layered materials is unclear. Recently, a new thermally activated volatile switching mechanism has been reported for a range of solution-processed 2D materials that can be exploited for artificial spiking neurons (Fig. 1i,j).[8] Here, the morphology of the 2D nanoflakes plays a vital role in producing non-linear behavior that can be used for high-order oscillator circuits. However, the lack of an effective selector has limited the integration of printed memristors in crossbar architectures (Fig. 1k,l). Recently, neuromorphic applications have also been proposed for 2D magnets, 2D charge density wave switches, and 2D moiré heterostructures, suggesting further opportunities in this space.

**Current and Future Challenges**
The main challenge facing vertical 2D memristors is competition with conventional metal oxide memristors that outperform their 2D counterparts in nearly all relevant metrics. Furthermore, wafer-scale 2D materials are generally polycrystalline, and spatial variations in grain boundaries are likely to lead to device-to-device variability, unlike the relatively high device-to-device homogeneity of amorphous metal oxide films.[3] This spatial inhomogeneity is further exacerbated by the inherent variability arising from stochastic switching that is common to all filamentary switches. While 2D memristors provide atomically thin channels, the lateral dimensions of metal lines are the more relevant scaling parameters for high-density crossbars, which also may be complicated by the finite grain sizes in 2D films. While single-crystal 2D flakes have also shown stable memristive switching arising from partially oxidized layers, wafer-scale growth of layered single crystals has not been shown. Thus, one immediate challenge in vertical memristors is to scale N x N crossbar arrays from N = 10 to N = 1000. Another key challenge is to integrate vertical memristors with a selector to avoid sneak current issues. A 2D transistor selector may be possible, although integration of a functional 1T1M crossbar has not yet been demonstrated.

Lateral memtransistors are faced with similar scaling challenges where the device footprint and operating power are not yet competitive with conventional vertical memristors. Since grain boundaries are believed to be essential for resistive switching in memtransistors, polycrystalline grain size likely dictates the ultimate scaling limits. Furthermore, since the operating mechanism of memtransistors relies upon the modulation of Schottky injection at the contacts, the operating voltage is not expected to scale linearly with channel length. Despite these challenges, the state-of-the-art complexity of lateral memtransistor crossbars (channel < 1 μm) and operating voltages (<1 V) approach that of vertical memristors (Fig. 1c,f).[3,4,6] Moreover, dual-gated lateral memtransistors achieve 1T1M functionality within the same device without requiring integration with another selector technology.[4,6] On the other hand, the switching speed of lateral memtransistors is significantly slower than vertical memristors, and gradual soft switching is likely to reduce the dynamic range of resistance under fast operating conditions.[2] Nevertheless, dual-gated ferroelectric and floating gate memtransistors have the potential to reduce the switching power and increase the switching speed.[2] Another challenge is integrating 2D NVM devices into circuits such as spiking neurons, activation circuits, and analog-to-digital converters (ADCs) for complete neural network chips.[1] Solution-processed 2D material devices also need to be scaled to sub-micron length scales for practical applications such as wearable electronics for off-grid classification and medical diagnostics.[7,8] In addition, the integration of printed NVM circuits with flexible logic circuits for full data processing has not been demonstrated. Overall, the grand challenges for 2D neuromorphic computing are centered on materials control and device engineering to achieve comparable metrics to conventional NVMs but with additional functionalities that yield improved efficiency in hardware computation.

**Advances in Science and Technology to Meet Challenges**
The last few years have seen significant advances in wafer-scale growth of 2D semiconductors and insulators that are directly relevant to the unique challenges of neuromorphic circuits.[3] Current growth advances are focused on achieving large grain sizes and minimizing lattice defects for wafer-scale uniformity of conventional transistor technology. For memristive devices using intrinsic defects, growth also needs to be optimized to yield small gain sizes (< 10 nm) and well-controlled defect densities.[1] Although wafer-scale 2D transistors have been used to realize neural network chips consisting of > 800 devices, this technology does not yet compete effectively with existing Si CMOS-based neural network chips. In this context, self-aligned vdW anti-ambipolar Gaussian heterojunction transistors have been shown to significantly simplify the circuit architecture of spiking neurons with a smaller number of elements than conventional CMOS circuits. These dual-gated Gaussian heterojunction transistors using mixed-dimensional nanomaterials can produce transfer functions resembling kernels such as Gaussian and sigmoidal kernels that are used in support vector machine hardware. A circuit with two such devices enables independent control of all parameters of mixed kernels that require close to 100 Si-based transistors in a conventional CMOS circuit, thus enabling highly energy-efficient machine learning.[9] These Gaussian transistors could also be integrated with a non-volatile memory (e.g., a floating gate or ferroelectric gate) to achieve Bayesian neural networks for predictions with confidence bounds. In terms of advances in fabrication, the self-aligned scheme also provides an opportunity for highly scaled lateral memtransistor crossbars. While efforts are underway to improve the performance of individual devices, the existing neuromorphic paradigms also need to be revisited to identify unique opportunities enabled by the unique characteristics of 2D devices. For example, recent algorithmic innovations in deep neural network architectures require higher-order processing where, along with inputs and model parameters (i.e., weights), the application context should also be considered in making predictions. For these higher-order neural networks, the additional gate electrode layer in dual-gated memtransistor crossbars presents a promising pathway to dynamic weight selection that mimics biological synapses.[10] In this manner, 2D neuromorphic computing has the potential to not only provide efficient hardware accelerators for machine learning algorithms but also realize emerging paradigms for bio-realistic neuronal hardware.[10]

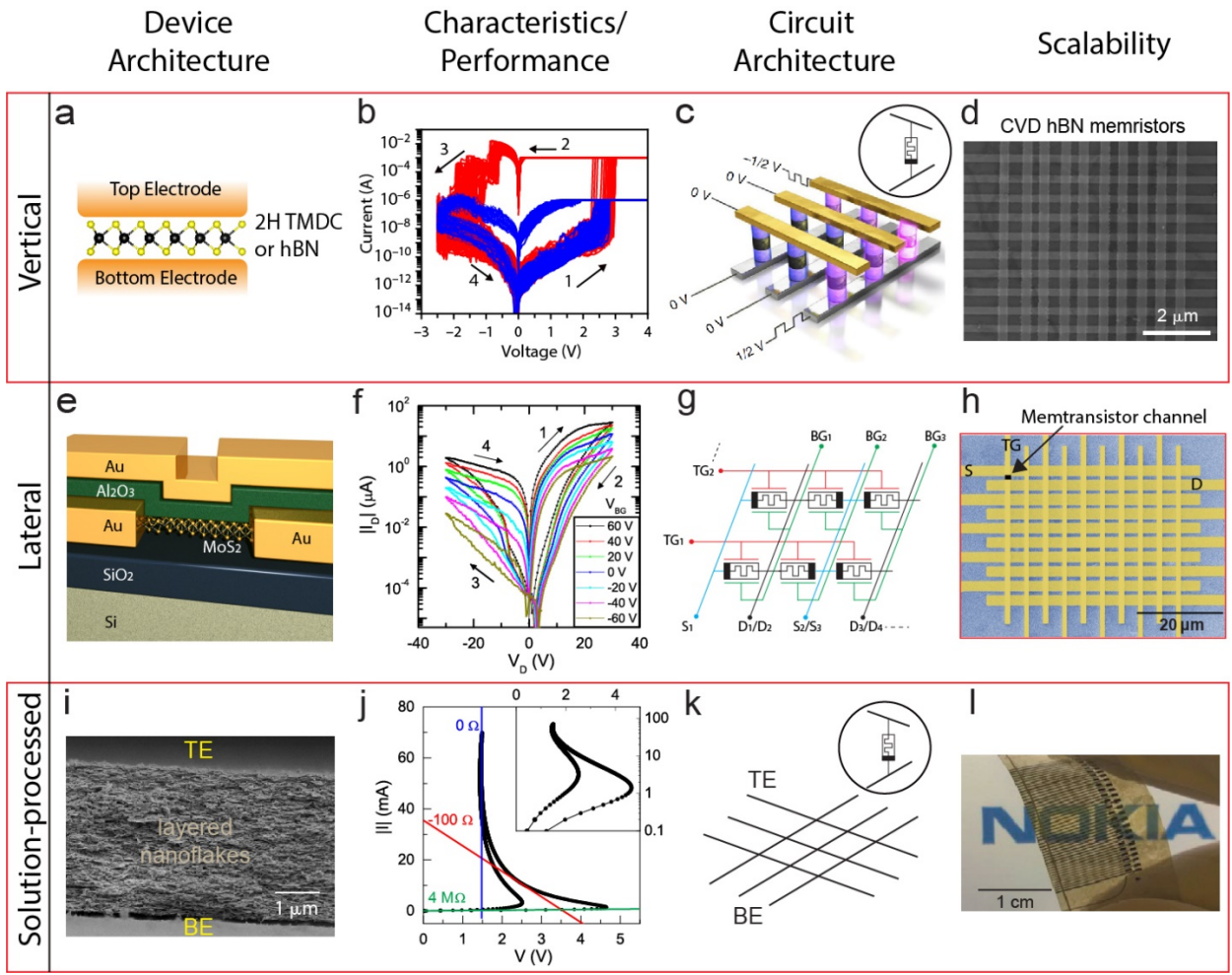

**Figure 1**. (a) Device architecture of a vertical memristor using layered materials such as TMDCs, hBN, and other insulators. (b) Typical current-voltage (I-V) characteristics of a vertical hBN memristor for two current compliances (red and blue curves). Arrows show the voltage sweep direction. (c) Crossbar architecture for vertical memristors where the desired node is selected by a V/2 biasing scheme. (d) Scanning electron microscopy (SEM) image of a 10 x 10 crossbar array of vertical memristors on a wafer-scale hBN film. (e,f) Device architecture and gate bias (VBG)- dependent I-V characteristics of a dual-gated lateral MoS2 memtransistor, respectively. (g,h) Architecture and SEM image (false color) of a dual-gated memtransistor crossbar array, respectively. (i,j) Cross-sectional SEM image and current-controlled I-V characteristics of a solution-processed MoS2 memristor, respectively. (k) Schematic of a 3 x 3 crossbar using memristors from solution-processed 2D materials. (l) Optical image of a 50 x 1 crossbar array of MoS2 memristors on printed Ag electrodes. (a) Reproduced with permission.[1] Copyright 2020, Springer Nature. (b,d) Reproduced with permission.[3] Copyright 2020, Springer Nature. (c) Reproduced with permission.[4] Copyright 2019, Springer Nature. (e-h) Reproduced with permission.[6] Copyright 2020, Wiley-VCH. (i,j) Reproduced with permission.[8] Copyright 2021, Wiley-VCH. (l) Reproduced with permission.[7] Copyright 2015, Springer Nature.

**Acknowledgments**

This work was primarily supported by the National Science Foundation (NSF) Grant Number CCF-2106964. V.K.S. and M.C.H. also acknowledge support from the Department of Energy (DOE) Threadwork Program under Grant Number 8J-30009-0032A and the Laboratory Directed Research and Development Program at Sandia National Laboratories (SNL). SNL is a multi-mission laboratory managed and operated by National Technology and Engineering Solutions of Sandia LLC, a wholly owned subsidiary of Honeywell International Inc. for the U.S. DOE National Nuclear Security Administration under contract DE-NA0003525. This paper describes objective technical results and analysis. Any subjective views or opinions that might be expressed in the paper do not necessarily represent the views of the U.S. DOE or the United States Government.

## 4.1- Computing with p-bits: A case study in the new era of electronics

Kerem Y. Camsari, Department of Electrical and Computer Engineering, University of California Santa Barbara, Santa Barbara, CA, USA (camsari@ece.ucsb.edu)

Peter L. McMahon, School of Applied and Engineering Physics, Cornell University, Ithaca, NY, USA (pmcmahon@cornell.edu)

Giovanni Finocchio, Department of Mathematical and Computer Sciences, Physical Sciences and Earth Sciences, University of Messina, Messina, Italy (giovanni.finocchio@unime.it)

Supriyo Datta, Elmore Family School of Electrical and Computer Engineering, Purdue University, West Lafayette, Indiana, USA (datta@purdue.edu)

**Status**

The slowing down of Moore's Law led to the emergence of an exciting new era of electronics. Following decades of continuous improvements in transistor technology, the new era is marked by blurred layers of abstraction in the computing stack, creative combinations of CMOS technology with emerging technologies and the development of domain-specific hardware and architectures. In this short piece, we describe probabilistic computing with p-bits, as a representative example of the many promising directions in the new era. We describe recent developments of p-bits; starting from their energy-efficient realization in different hardware substrates, their physics-inspired and parallel architectures, all the way up to their use in high-level algorithms and applications. A recurring theme in this piece is that of *co-design*; where algorithms and applications are modified to naturally conform to the underlying physics of hardware. Along with domain-specificity, co-design will play an increasingly important role in the new era which will be marked by various CMOS + X type heterogeneous architectures.

**Current and Future Challenges**

In a celebrated talk delivered at a conference in May 1981, Richard Feynman introduced the first clear vision of a quantum computer [1]. The main idea of Feynman's talk, appropriately titled "Simulating Physics with Computers", can be summarized by the credo "Let physics do the computing." In other words, simulating physical phenomena is efficient when the simulating "computer" itself is made of the building blocks it is trying to simulate. This profound connection between physics and computing Feynman emphasized has since been used to develop quantum computers built out of quantum mechanical bits. This part of the story is very well-known and often discussed, see, for example, "Quantum Computing: 40 Years Later" by John Preskill for more details [1]. What is less appreciated is that before getting on to quantum computing, Feynman talked about a vision of a probabilistic computer with essentially the same idea: a probabilistic *Nature* should be efficiently simulated by a machine that itself makes probabilistic decisions. In a few lines, Feynman laid out the main idea that is used in many probabilistic models today: in an interacting system with many degrees of freedom, if we need to compute correlations between small parts of the system, all we have to do is observe those parts. What is otherwise an intractable summation over the exponentially large "rest of the system states" then becomes approximately tractable[1].

Driven by the nearing end of Moore's Law, a few years ago, we took Feynman's vision a step further. Intrigued by the large degrees of inherent noise in magnetic nanodevices, we imagined a truly probabilistic computer down to its most basic building block. Early work involved using the probabilistic switching behavior of stable magnets [2], but gradually the temporal noise of low barrier nanomagnets (LBM) became a more natural choice. LBMs offered the possibility of a compact realization of the basic building block of a probabilistic computer, which we named the "p-bit", and we experimentally demonstrated a prototype "p-computer" shortly after [3].

---

[1] A technical note for the expert: the observed correlation is *still* an approximation unless we observe the system with time T → ∞ which amounts to computing the intractable sum exactly, for an ergodic system.

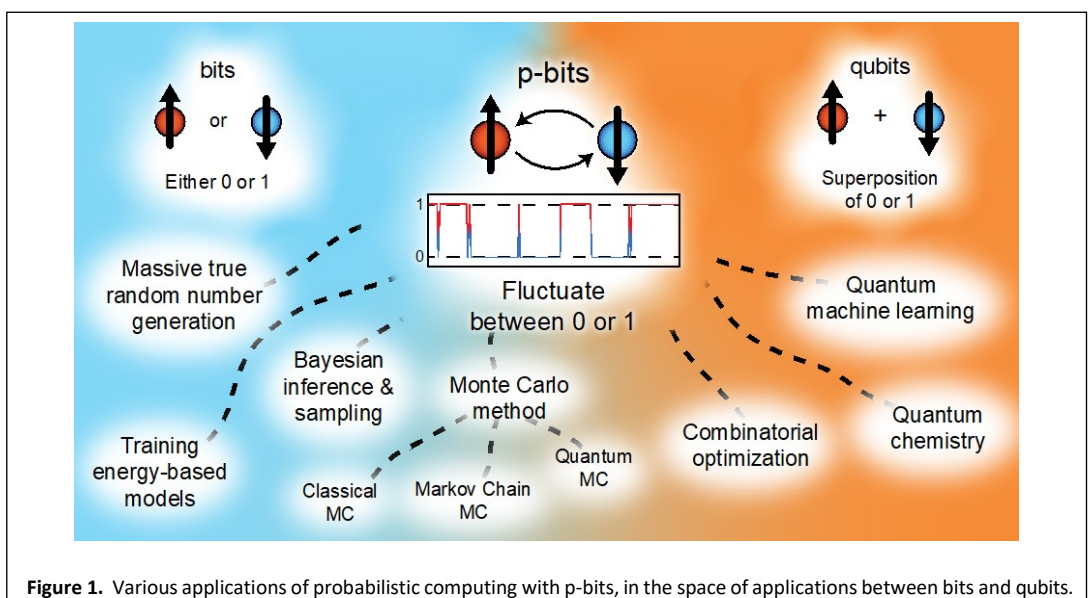

**Figure 1.** Various applications of probabilistic computing with p-bits, in the space of applications between bits and qubits.

The ubiquitous nature of probabilistic methods and randomized algorithms allows p-bits to be applied to a broad range of applications (FIG. 1). Examples include massively parallel true random number generation, solving combinatorial optimization problems using powerful algorithms such as simulated annealing and parallel tempering, probabilistic sampling for Bayesian inference and learning, training energy-based and variational classical and quantum models, accelerating Monte Carlo (MC), Markov Chain Monte Carlo (MCMC), Quantum Monte Carlo methods, computational biology and protein folding among others. The wide-ranging application space for p-bits makes them potential candidates for domain-specific computing, with overlapping applications envisioned for near-term quantum computers, particularly for Machine Learning and AI applications.

**Advances in Science and Technology to Meet Challenges**

*Device-circuit co-design of p-bits.* In essence, a p-bit is the abstraction of a tunable Bernoulli variable with many different physical implementations. p-bits are closely related to the basic unit of Boltzmann Machines, the binary stochastic neuron, pioneered by Hinton and Sejnowski [4]. The p-bit, when defined as a mathematical abstraction [3], has a much wider range of applications than just Boltzmann Machines (FIG. 1). Therefore, finding the most energy-efficient, technologically scalable, and robust p-bit is an active area of research. In addition to magnetic p-bits with stochastic magnetic tunnel junctions, stochastic resistors (e.g., diffusive memristors, perovskite nickelates), diodes (e.g., Zener, single photon avalanche) and even analog or digital CMOS (e.g., RTN in silicon transistors, LFSRs) can make compact and energy-efficient p-bits. Similarly, connecting p-bits to one another can be achieved in many ways: resistive (or capacitive) crossbar arrays, and digital or mixed-signal CMOS-based interconnections are examples, to name a few. Key metrics in designing good p-bits are the energy (E) and delay ($\tau$) to produce a random bit. Like novel switches, minimizing the energy-delay product of a truly random bit with the minimum area footprint guides the development of novel p-bits. Exciting new experimental developments with stochastic magnetic tunnel junctions [5] have shown that $\tau$ can be a few nanoseconds or less. Combining various possible MTJ designs (e.g., in-plane, circular disk, perpendicular, double-free-layer) with CMOS transistors by deliberate co-design, energy-efficient circuits with E < 1 fJ/rng could be obtained in monolithically integrated p-computers with tens of millions of p-bits. The key advantage of a nanodevice-based p-bit comes from the large savings in energy, area, and the quality of randomness over digital CMOS. Even when compared to low-quality pseudo-random number generators in CMOS, an MTJ-based p-bit is at least 1000X smaller in area and 100X smaller in energy to produce a random bit. The compactness in the area and energy efficiency opens up the potential for a high degree of scalability with MTJ-based p-bits, beyond what is accessible with present-day technology.

*Architecture-algorithm co-design of p-bits.* A key step in mapping algorithms to hardware is to find an efficient architecture co-designed with the algorithm. Designing probabilistic computers starting from single devices to systems allows imagining completely new architectures with suitably modified algorithms. To illustrate this point, consider a simple MC algorithm for calculating the number ``π'': Imagine a square with a circle in the center and divide each side of the square into $2^{20}$ segments. Then, use a (20 + 20)-bit RNG to generate a random coordinate (x, y). Calculate an output s ∈ {0, 1} indicating whether the random coordinate (x, y) lies inside the circle or not (s = $\mathbb{1}[x^2 + y^2 < 1]$). Perform *N* trials and obtain an average to estimate π. This is clearly a parallelizable algorithm but not trivially: avoiding significant delays while calculating the sum of squares requires a carefully pipelined architecture such that a new sample is obtained at every clock cycle [6]. Similarly, accelerating Markov Chain MC algorithms requires deliberate designs since for directly connected p-bits, parallel updates are not allowed. One such design is exploiting the idea that not directly connected (conditionally independent) p-bits can be updated in parallel. This allows for reaching high levels of parallelism if the connecting graphs are sparse and can be divided into large segments that are updated in parallel [13]. More intriguingly, p-computers can have entirely asynchronous architectures where each p-bit updates with its randomly ticking internal clock in physics-inspired, massively parallel architectures. Preliminary results indicate the promise of such architectures; however, our main point is to illustrate the wide range of possibilities that exist for architecture-algorithm co-design.

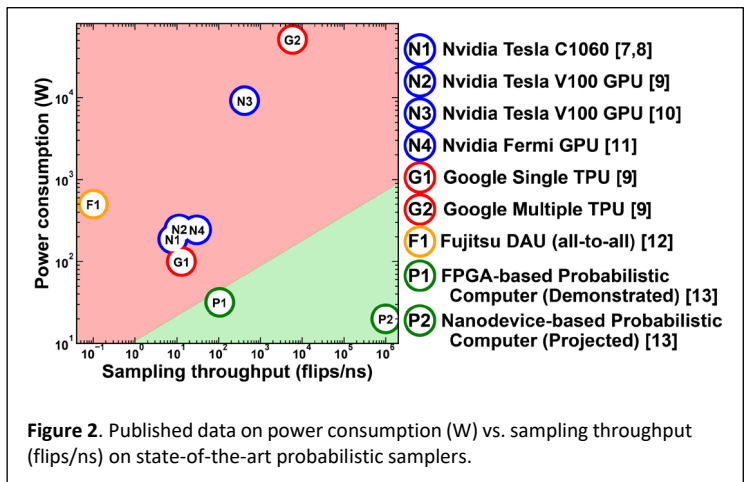

**Figure 2**. Published data on power consumption (W) vs. sampling throughput (flips/ns) on state-of-the-art probabilistic samplers.

*Benchmarking probabilistic computers.* Virtually all applications shown in FIG. 1 benefit from one key metric in probabilistic computers, namely the sampling throughput or flips per second [7-13]. Sampling throughput is commonly reported in specially designed probabilistic samplers and FIG. 2 shows power consumption vs. sampling throughput for highly optimized implementations. GPU and TPUs often use highly regular graphs (typically 2D nearest neighbor grids) to achieve scalability in their architectures. Others, such as Fujitsu's digital annealer use all-to-all connected graphs, taking fewer samples per second but compensating for this by means of powerful algorithms such as parallel tempering and population annealing [12]. For example, the Google TPU can take more than 5,000 flips/ns but at the expense of 50,000 W power dissipation to achieve this feat! On the other hand, FPGA-based p-computers take 100 flips/ns using around 20 W. But more importantly, projections based on nanodevice-based p-bits indicate the possibility for 1,000,000 flips /ns at only 20 W of power! This number can be reached in designs where each p-bit dissipates 20 μW with a million of them flipping every nanosecond in an asynchronous setup. All of these pieces have been individually demonstrated: the magnetic memory industry has scaled MTJs up to billion-bit densities and stochastic magnets have been shown to fluctuate every nanosecond. The potential for growth and acceleration by p-bits seems highly promising if challenges for integration and co-design can be surmounted in the future.

**Concluding Remarks**

Similar in spirit to many other promising domain-specific computing paradigms, there are several important areas requiring further attention. From the physics end, identifying the best possible mixed-signal p-bit design is still a work in progress. Controlling device-to-device variations or overcoming them through algorithm hardware co-design is also critical. From the systems side, identifying and adopting powerful algorithms and applications conforming to p-bits requires expert algorithmic understanding. Optimizing the necessary bit precision, and design modes (synchronous vs. asynchronous) with the right architecture, while being amenable to monolithic integration all require a concerted effort and the widest possible expertise in the computing stack. Overall, this exciting, full-stack research program is simply one example of a powerfully emerging trend in the new era of electronics where domain-specific hardware and architectures will play an increasingly important role.

**Acknowledgment**

KYC is grateful to N. A. Aadit and A. Grimaldi for assistance in the preparation of the figures. KYC acknowledges support through the National Science Foundation (CAREER Award), Samsung GRO program, and Office of Naval Research Young Investigator Program. The work of GF was supported under the project PRIN 2020LWPKH7 funded by the Italian Ministry of University and Research.

## 4.2- Quantum computers with spin-based qubits in silicon

Belita Koiller, Instituto de Física, Universidade Federal do Rio de Janeiro, 21941-972 Rio de Janeiro, RJ, Brasil (bk@if.ufrj.br)
Gabriel H. Aguilar, Instituto de Física, Universidade Federal do Rio de Janeiro, 21941-972 Rio de Janeiro, RJ, Brasil (gabo@if.ufrj.br)
Guilherme P. Temporão, Center for Telecommunications Studies, Pontifical Catholic University of Rio de Janeiro, 22451-900 Rio de Janeiro, RJ, Brasil (temporao@puc-rio.br)

**Status**

Universal Quantum Computers (QCs) can potentially solve open problems that are not only relevant to science and technology but also likely to assist current social demands and expectations, e.g. regarding food supplies and environment issues.

The formalism for quantum information processing is substantially simplified by the following result: A universal set of gates, consisting of all one-qubit quantum gates and a single two-qubit gate, e.g. the controlled-NOT (C-NOT) gate, may be combined to perform any logic operation on arbitrarily many qubits (thus pointing a clear path towards universal QC). In addition, few requirements must be fulfilled: 1) Proposals should eventually provide a prototype performing any operation compatible with universality; 2) The fidelity of one- and two-qubit operations must be above 99%; and 3) The QC architecture is expected to fit within a few centimeters chip.

Among a few existing quantum technologies compatible with these criteria are electron spins in semiconductors, which we briefly review here. Semiconductor-based QC started as a promising candidate for implementation of QC [1], by confining electrons within electrostatically defined quantum wells in a GaAs/AlAs interface 2DEG. Changing the height of the barrier separating two electrons in neighboring wells, 2-qubit quantum operations can be performed (top Fig.1a). These ideas were adapted to qubits defined by spins of electrons bound to shallow donors in Si [2]. Even though the original proposal suggested impurity nuclear spins-1/2 as qubits, such as P in Si, it was soon recognized that nuclear spins are much too isolated from the environment for fast, efficient control by external fields. A more promising route considered the spin of the extra electron in P in a Si host, which is loosely bound to the donor (lower Fig.1a). The long-lived P nuclear spins have been considered appropriate for quantum memories [3].

Two decades after the first proposals, silicon qubits still hold the status of a great promise. They have been shown to exhibit long enough coherence times, high-fidelity gates, fast operation capabilities and a huge potential for a scalable solution. Indeed, the search for a scalable and universal silicon-based quantum computer has so far attracted full attention from active research groups in academic institutions, major industrial labs and start-up companies. Moreover, spin-based QC uses the same technology adopted in fabrication of transistor-based electronics, benefiting from the established know-how and huge investment in the silicon technology. Notwithstanding all advantages, there are many challenges yet to be overcome, as described in the next sessions.

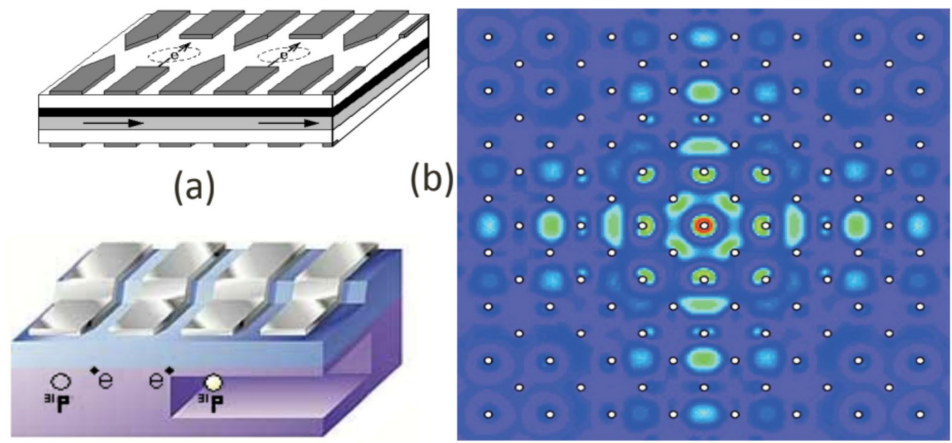

**Figure 1.** (a) Schematic representation of early (1998) QC architectures with spin qubits in semiconductors : - upper scheme represents Loss-DiVincenzo's proposal on GaAs (qubits bound to quantum wells [1]); bottom scheme represents Kane's proposal involving P substitutional donors in Si [2]. (b) Electron probability density around a P substitutional donor on the (001) plane of bulk Si for the donor ground state. The white dots give the in-plane atomic sites. Obtained from ref. [4].

**Current and Future Challenges**

Silicon qubits are currently undergoing a major transition, from lab prototypes featuring a few qubits to massive scale production. Scalability is a key ingredient in demand for progress in any QC platform. Recent progress reported by a joint Intel – QuTech partnership [5] demonstrated a fully optical lithography technique - similar to those in use in current integrated circuit fabrication methods - in order to obtain more than 10,000 arrays of quantum dot (QD) spin qubits on a single 300-mm wafer (Fig. 2). This leads to a device yield above 98% and good QD uniformity, with a normalized standard deviation in the gate threshold voltages around 7%. The spin relaxation and dephasing times $T_1 \sim 1s$ and $T_2^* \sim 20\mu s$ - which can be improved to the 3 ms range, or even higher, by dynamical decoupling. Such indicators stress the huge potential of QC using QDs in silicon.

There are many challenges, however, remaining to be addressed. Spin qubits require individual control, involving a connection to a classical auxiliary electronic device. This creates serious difficulties to implement the required individual wiring when the number of qubits is in the order of magnitude of millions. Qubit readout is also a sensitive aspect, as not only the readout bandwidth must be much faster than the spin decoherence time but the readout of a single qubit must not interfere with the neighboring qubits [6]. Concerning donor spins in Si, fabrication and control of multi-qubit arrays are among the most critical limitations.

Remaining challenges concern adapting the large-scale manufacturing process for two-dimensional spin arrays, which are important for two main reasons: being able to implement surface codes for robust quantum error correction and optimizing the efficiency of cryogenic cooling of the circuit [7]. Moreover, two-qubit gates still need to be fully implemented and characterized; fault-tolerant quantum computing requires two-qubit fidelities of at least 99%, a threshold which has been overcome very recently for QDs in silicon [8]. In fact, circuit characterization and benchmarking - including defining figures of merit for assessing the quality of the QC along all the manufacturing steps, which are very different from the classical counterpart - are among the main challenges that need to be tackled to keep any expectations of achieving a QC with millions of qubits.

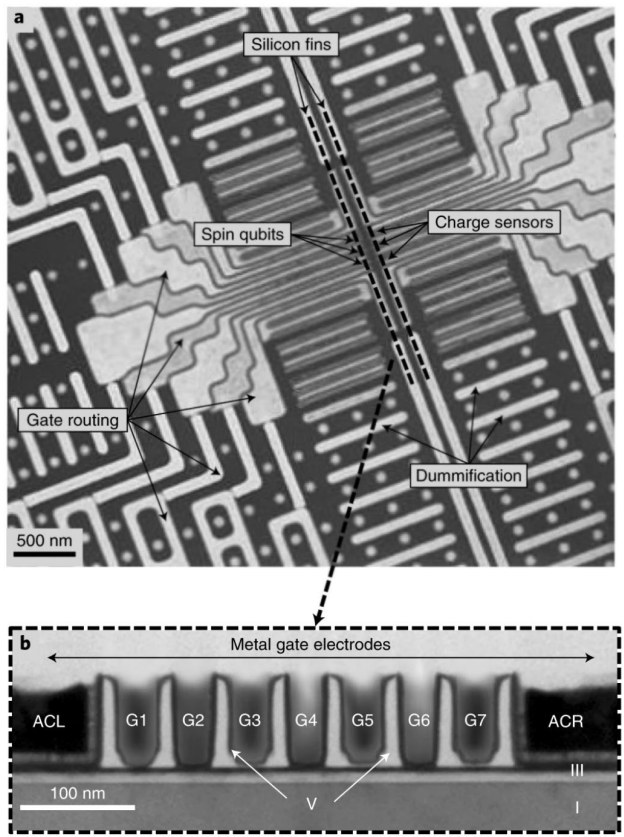

**Figure 2**. (a) Electron microscopy image of an industrially fabricated QD device by Intel and QuTech, showing two parallel silicon fins, one hosting the qubits (left) and the other hosting the sensing QDs (right). Gate routing and dummification (required for maintaining a constant metal density on the surface) are also clearly shown. (b) Image along a Si fin of a QD array showing seven metallic silver gates (G1-G7) between two accumulation gates (ACL and ACR). Figure adapted from ref. [6].

**Advances in Science and Technology to Meet Challenges**
In order to scale up the spin system to reach a fully-fledged QC, we need to avoid the heat generated by every qubit added. This is especially critical in the case of superconducting devices, limiting the number of qubits per dilution-refrigerator. This is not the case for spin qubits, which can operate in an environment of up to 2K [9]. Therefore, the cost on the refrigeration can be reduced significantly in spin QCs, and the scaling-up is likely also be favored. The answer is unknown, as current simulation software employed by the electronics industry is not designed to properly deal with low temperature behaviour. Promising candidates for overcoming this limitation include Contact Block Reduction-based Quantum Computer Aided Design (QCAD) [10].

Problems with spin initialization and readout are usually mitigated by employing a conversion spin-charge. This can be implemented by using a reservoir or by Pauli spin blockade (reservoir-independent). Dispersive read-out has also been considered for single and few electrons as well as silicon nanowire transistors. However, the spin blockade has been advantageous so far because it allows higher readout fidelities and lower qubit operation frequency (~ 1GHz), opening the possibility of reading-out a qubit array in a time smaller than the decoherence time of the single qubit (millisecond range) [11].

To overcome the engineering challenge of simultaneous addressing many qubits in a large-scale spin-based QC, electron-spin-resonance techniques in conjunction with Kane's proposal ideas may be utilized. This technique could include a three-dimensional dielectric resonator that acts as a single global source that can deliver multiple control signals to the qubits. Recent advances show that such resonators, constructed from potassium tantalate (KTO), could be manufactured within a compact surface area of $0.7 \times 0.55$ mm$^2$, allowing its integration to nanoelectronic circuits and performing large-scale control over millions of spin qubits [12].

Finally, high volume fabrication of spin qubits requires a proportional capacity for characterization and tests. This is currently delaying the evolution of spin-based QC, due to lengthy tests in low-temperature environments, which require cooling down the devices in dilution refrigerators. This limitation could be avoided by a thorough characterization of the correlation between classical semiconductor device metrics (such as mobility) and spin qubit performances, but no definitive results have been obtained so far. Alternative characterization procedures, possibly performed at higher temperatures, are still missing, but some effort concerning this huge challenge can already be found in the literature [13].

**Concluding Remarks**

Scalability is one of the most important challenges that any architecture needs to achieve in order to obtain a quantum computer that is able to solve relevant problems. Spin qubits are possibly the most suited candidates for building a universal QC, especially because of its potential scalability which has already been demonstrated experimentally. Other ongoing debates relevant for the future development of spin-based QC include: finding the most effective way of encoding quantum information in silicon (e.g. electrons vs. holes, quantum dots vs. donors, Loss-DiVincenzo vs. Singlet-Triplet qubits), which figures of merit need to be measured for proper device characterization (and how they should be measured), the best way to perform spin readout, which quantum error correction codes should be employed, and so on. In any case, QC is an inherent multidisciplinary field, and as such, it is expected that physicists, electrical engineers and computer scientists participate in discussions, working together towards a large-scale QC.

**Acknowledgements**

The authors thank André Saraiva for the interesting and very fruitful conversations. The authors acknowledge financial support from the Brazilian agencies CNPq (PQ Grants No. 307058/2017-4, INCT-IQ 246569/2014-0 and 307910/2019-9). G.H.A. and G.P.T. also acknowledge FAPERJ (Grants No. 210.069/2020 and 211.094/2019, respectively) and FAPESP (Grant No. 2021/96774-4).

## 5.1- Computing with Dynamical Systems

Davi R. Rodrigues, Department of Electrical and Information Engineering, Politecnico di Bari, Italy (davi.rodrigues@poliba.it)
Satoshi Sunada, Faculty of Mechanical Engineering, Institute of Science and Engineering, Kanazawa University, Japan (sunada@se.kanazawa-u.ac.jp)
Karin Everschor-Sitte, Faculty of Physics and Center for Nanointegration Duisburg-Essen (CENIDE) University of Duisburg-Essen, Germany (karin.everschor-sitte@uni-due.de)

**Status**

Dynamical systems include particles or ensembles whose behavior over time is accurately described by a set of equations, such as electrical circuits, magnetic systems, chaotic systems, and spin glasses. Interestingly, computational applications can exploit the intricate behavior of these ubiquitous systems. This is achieved by deliberately engineering the system's spatial and temporal dynamics to emulate a particular computational task, as illustrated in Fig. 1. The use of dynamical systems for computing applications dates to analog computers, which were replaced by digital computers in the mid-1990s. In recent decades, digital computers have offered greater versatility, less susceptibility to failure, and better scalability, made possible by the rapid development of transistors [1]. Dynamical systems are used in a plethora of computing applications, including random-Boolean networks, Ising machines (see sections 5.2 and 5.3), memcomputing (see section 5.4), and neuromorphic computing (see sections 1.1, 1.2, 1.3, and 1.4 for example). In the latter context, dynamic systems offer several advantages over digitized circuits and are thus making a comeback. Dynamical systems can be designed to have the required structural similarities of neural networks, such as hierarchy, approximate symmetries, memory, redundancy, and nonlinearity [2]. Therefore, they provide a natural hardware implementation of neural networks that overcomes the von Neumann bottleneck with much lower power consumption and higher scalability than transistor-based digital technology, which only artificially emulates the required properties. In addition, analog information processing in dynamical systems allows sensor signals to be processed directly, providing energy-efficient and low-latency processing. The use of dynamical systems for computing has been facilitated by recent advances in materials science, such as breakthroughs in photonics and spintronics, which have enabled the development of low-power proof-of-concept devices compatible with CMOS technology. Figure 2 shows examples of proposed nanoscale dynamic systems for analog computing. These systems are designed to operate at extremely small dimensions and exploit the unique properties of nanomaterials and nanoscale phenomena to perform analog computations. Recent proposals have shown that dynamical systems can be used to emulate neuron synapses and firing for synaptic neural networks [3], as well as to perform weight computations or completely replace hidden layers in a neural network [2,4–7]. Two promising applications that consolidate the use of dynamical systems for spatio-temporal pattern recognition are reservoir computing [5,7] and physical neural networks [2–4,6,8]. While both computational paradigms allow for learning and extracting patterns from data, they rely on different learning schemes. In reservoir computing, training is performed only at the output level and thus requires a sufficiently complex response of the physical reservoir to discriminate minor variations in the input. In contrast, physical neural networks train directly on the physical system, often requiring complex learning algorithms. In general, a physical realization of a reservoir computer benefits from a system with highly nonlinear dynamics and short-term memory. In contrast, a physical neural network requires a well-modeled nonlinear system with controllable dynamical parameters.

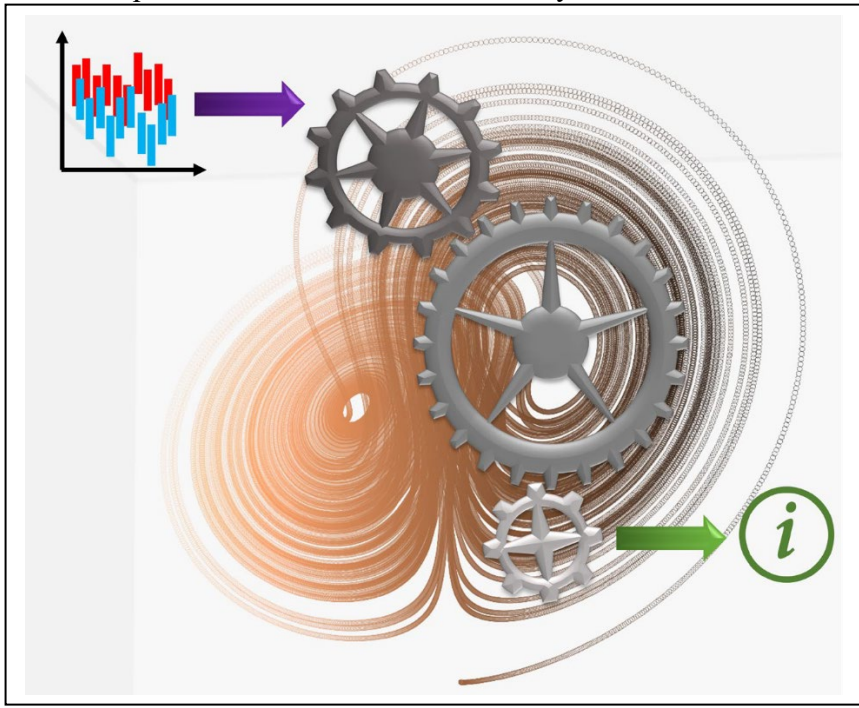

**Figure 1.** Computing with dynamical systems: A dynamical system computes based on the input signal and generates information as output. The complex and nonlinear dynamics of the system generates a final state, corresponding to the output, that depends on the initial state of the system, corresponding to the input, and tunable external perturbations. By properly designing the dynamical system, it is possible to tune the final state to be a desired function of the initial state and external perturbations. This allows fast and efficient computations.

**Current and Future Challenges**

Computing with dynamical systems typically implies the encoding of computational tasks into the functional response of the device. Thus, current and future challenges in the field are related to designing devices with tailored functionalities and ensuring reliable encoding of information in terms of inputs and outputs of the dynamical system. While dynamical systems can be tailored to perform complex computations with higher efficiency compared to digital computers, they are often single-purpose devices with analog outputs. Designing dynamical systems as multi-purpose devices is a key challenge. In addition, training and learning strategies must become more efficient when computing with dynamical systems. In the future, they must go beyond simple supervised learning models.

Furthermore, a significant challenge for the commercial implementation and widespread use of dynamical systems-based computers, for example, in the Internet of Things and Industry 4.0 applications, as well as in real-time computing, is to produce devices that are scalable, inexpensive, and significantly more efficient than CMOS technology. At least on intermediate time scales, it is also necessary to be able to integrate them into the market-dominating CMOS technology. This integration requires significant efforts in materials science and device engineering. Although the number of proof-of-concept devices is growing rapidly, most still face significant obstacles in their large-scale realization and production.

A fundamental question concerns the general design of the device. It can be realized either with an assembled network of individual simple components or directly with a large complex system. Both strategies have advantages and disadvantages and are likely to be used for different types of applications. For example, a large complex system is often less tunable. However, it can avoid the challenge of creating a highly dense interconnected network of individual components, which only within the network structure allows for high computational performance.

Another major challenge is that dynamical systems computers are not fully in-situ. For example, the input data must be pre-processed by an external device to generate a significant nonlinear response of the dynamical system, which typically responds only on specific time and length scales [9]. Similarly, dynamical systems often provide a continuous set of outputs that must be interpreted or even learned (as in reservoir computing) by an external computer. An autonomous device with a reliable map between the input, the functional response of the device, and the correct output is still missing.

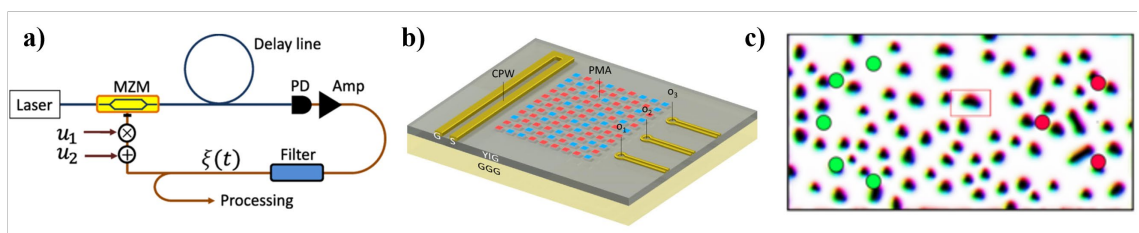

**Figure 2.** Examples of computing with dynamical systems. a) Optoelectronic delay system. Reproduced from Ref. [10], b) Nanomagnet-based spinwave scatterer. Reproduced from Ref. [4], c) Reservoir Computing based on Skyrmion fabrics. Reproduced from Ref. [5].

**Advances in Science and Technology to Meet Challenges**

The field of computing with dynamic systems is steadily growing with many proposals to address the challenges, ranging from the development of novel computational algorithms [2,10–12] to efficient manufacturing techniques. The main goal is to develop an efficient algorithm-and-hardware codesign to fully exploit dynamical features assisted by state-of-the- art nanotechnologies. The interdisciplinary approach is of crucial importance and ensures knowledge transfer in particular within computer science, mathematics, biology and physics. A key example that connects all disciplines is the development of novel brain-inspired algorithms that are then transferred to computing in materials and devices.

The development of techniques to quantify, tailor and exploit the nonlinear response and short-term memory of dynamical systems provides means to reduce the complexity and energy cost of conventional learning algorithms and to approximate the biologically inspired behavior of the brain, which is optimized by evolution over millions of years. Structured material studies based on machine learning are used to achieve advances in material properties. A particularly successful example in material science is the recent progress in manufacturing heterostructures from 2d materials and metamaterials [13]. These multiphysics systems with potentially different natural timescales offer a variety of physical properties that make them attractive as highly scalable and tunable platforms for novel unconventional computing schemes.

In addition, there are many approaches in device development to implement neuromorphic functionalities in dynamical systems including targeted studies to improve device topologies and device designs. The development of scalable multifunctional systems is crucial for the various design concepts, such as considering a network of coupled individual units or a single large complex system. Current proposals target the use of scalable systems based on time- delay structures [7,10], spatial parallelism [8] of photonic systems, spin nano-oscillators [9], and patterned magnetic samples [4,5]. Furthermore, there are focused efforts to integrate as many functionalities as possible directly in-situ into the device. These include the implementation of new learning algorithms based on physical properties [10,12] and the engineering of the functional response of the systems [6]. Fully autonomous devices, that can perform both calculations and learning, however, will still demand great efforts.

**Concluding Remarks**
Computing with dynamical systems is an exciting field of research, which is experiencing a revival primarily driven by the significant advances in neuromorphic computing. The demand for efficient and scalable hardware implementations of neuromorphic systems, which can naturally be emulated in dynamic systems, brings the field from a niche to the forefront of research. Rapidly advancing significant developments in material science promise low-cost, easy-to-manufacture, and highly efficient task-oriented devices in the future. The ever-expanding capabilities to directly manipulate physics at the nanoscale and ultra-high frequencies expand the possibility of employing physical principles for novel algorithms and computational schemes that fully embody the brain's functionalities.
In summary, computing with dynamical systems has the potential to overcome the high-power consumption as well as the scalability limitations imposed by current CMOS technology and to actively shape the development of Industry 4.0 and the Internet of Things.

**Acknowledgements**
We thank Jake Love for discussions. DRR acknowledges funding from the Ministerio dell'Università e della Ricerca, Decreto Ministeriale n. 1062 del 10/08/2021 (PON Ricerca e Innovazione). SS acknowledges supports from JSPS KAKENHI (20H04255) and JST PRESTO (JPMJPR19M4). KES acknowledges funding from the German Research Foundation (DFG) Project No. 320163632 and the Emergent AI Center funded by the Carl-Zeiss-Stiftung.

## 5.2- Simulated Bifurcation

Kosuke Tatsumura (kosuke.tatsumura@toshiba.co.jp)
Hayato Goto (hayato1.goto@toshiba.co.jp)
Toshiba Corporation.

**Status**

Simulated bifurcation (SB) [1] is a quantum-inspired heuristic algorithm for finding the exact or approximate ground states of Ising spin models and is expected to be useful for various practical combinatorial optimization. Many combinatorial optimization problems are classified as non-deterministic polynomial-time (NP)-hard, where the computational complexity scales exponentially with the problem size, and can be converted to Ising problems [2]. Special-purpose hardware devices for solving Ising problems are called Ising machines, including SB-based machines.

The SB algorithm was found as a classical counterpart of bifurcation-based adiabatic quantum computation with a nonlinear oscillator network [1]. In SB, we numerically simulate the adiabatic evolution of a classical Hamiltonian dynamical system (a nonlinear oscillator network) with bifurcations (Fig. 1). Two branches of a bifurcation in each nonlinear oscillator represent two states of each Ising spin. The operational mechanism of SB is based on an adiabatic and ergodic search (Fig. 1) [1]. Recently two other variants of SB called the ballistic simulated bifurcation (bSB) and the discrete simulated bifurcation (dSB) [3] have been proposed and demonstrated to outperform the original adiabatic SB (aSB) in terms of both speed and solution accuracy. These algorithms exploit new effects, such as a quasi-quantum tunneling effect [3].

The SB algorithms are highly parallelizable and thus can be accelerated with massively parallel processors such as FPGAs (field-programmable gate arrays) and GPUs (graphics processing units) [1,3,4,5]. SB allows us to simultaneously update $N$ coupled-oscillator variables for $N$-spin problems at each time step. This is in contrast to simulated annealing (SA, a conventional heuristic algorithm), which involves sequential updates of spins, with simultaneous updates allowed only for isolated spins. For $N$-spin Ising problems with full connectivity, the maximum numbers of parallelizable operations in SB and SA are, respectively, $N^2$ and $N$ [4,5]. Custom-circuit implementations of SB [1,4] have demonstrated a higher degree of computational parallelism than the problem size $N$ (8,192 parallel processing elements for 2,048-spin Ising problems).

Various Ising machines based on different principles such as SA, quantum annealing, and dynamical-system evolution have been implemented with a variety of technologies including superconducting circuits, optics, emerging nanodevices, parallel digital processors, etc [2]. SB-based machines have been evaluated for various benchmark problems and compared with other Ising machines [2,3], demonstrated to be highly competitive, especially showing the highest performance for Ising problems with full connectivity [known as the Sherrington-Kirkpatrick (SK) model].

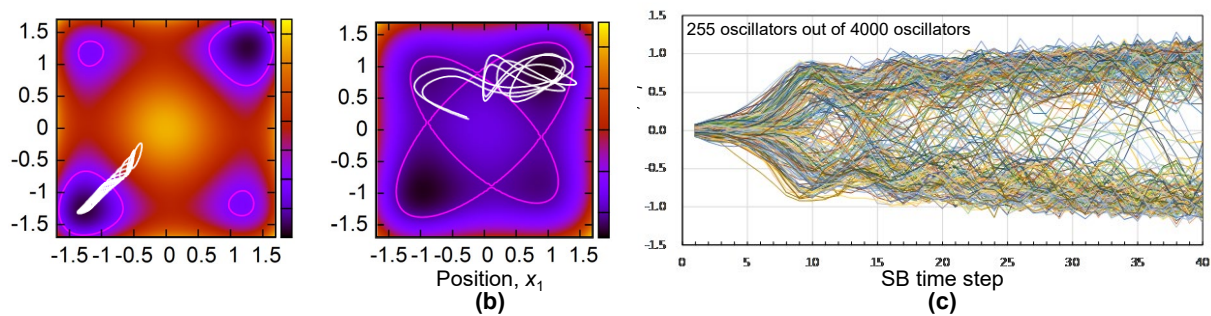

**Figure 1.** Dynamics in simulated bifurcation. (a) and (b) show trajectories indicating adiabatic and ergodic searches, respectively, for a two-spin problem ($N$=2) [1]. (c) Time evolution of oscillators exhibiting bifurcations ($N$=4000) [4].

**Current and Future Challenges**

SB is theoretically new (published in 2019 [1]) and there are many challenges and opportunities for further enhancement and wider applicability.

While quantum adiabatic optimization is based on the quantum adiabatic theorem [1], the operational mechanism of SB (adiabatic and ergodic search), which implies the classical adiabatic theorem, has been only empirically understood [1]. The mathematically rigorous proof of the operational principle of SB as well as the convergence property have been left for future work [1]. Potential theoretical studies include extending SB to polynomial unconstrained binary optimization (PUBO), relating SB with nonequilibrium statistical mechanics, and combining SB with techniques for complex constraints. Since the comparison between various Ising machines in terms of performance depends on problem instances, figure-of-merits and physical implantations [2,3], comprehensive and systematic comparisons should be continued.

Building larger Ising machines while avoiding speed degradation is challenging. In SB, the matrix-vector multiplication (MM) of the coupling matrix $J$ and the position vector $x$ of nonlinear oscillators (many-body interaction) is the most computationally intensive part [4]. To process the MM part in a massively parallel fashion, we have to prepare many processing elements (multiply-accumulators) and supply the $J$ and $x$ data to the processing elements at a sufficient transfer rate (the transfer rate needed increases with increasing the processing elements). As an example, the FPGA implementations of SB [1,3,4] were equipped with optimized memory subsystems to supply the $J$ and $x$ data by using on-chip memory (having larger bandwidth than external memory) and thus were allowed to fully utilize the computation resources in the time domain. However, the machine size (maximum problem size) of such a single-chip implementation is limited by the on-chip memory resource. Hence enlarging the machine size while fully utilizing the computation resources is of importance. The possible two approaches are scale-up (making a chip larger or denser) and scale-out (increasing the number of networked chips). The SB-based machines would benefit from emerging nanodevices for processing, memory, and communication in conjunction with in-memory computing, stochastic computing, and cluster computing architectures.

By implementing not only SB processing circuits but also interface/control circuits on a single chip, we can shorten the system-wide latency, enabling real-time systems based on combinatorial optimization that make the optimal responses to ever-changing situations. SB-based systems are thus expected to realize innovative applications.

**Advances in Science and Technology to Meet Challenges**

SB has been receiving increasing attention because of both the high performance and high practicability. Several advances in theoretical extension [6], custom-circuit architecture [5], and applications [7-10] are as follows.

Kanao *et al.* introduced a heating process to the SB Hamiltonian dynamics to assist the system during the search to escape from local minima, leading to improved performance [6]. This method was inspired by the Nosé-Hoover method for simulating Hamiltonian dynamics at finite temperature. The heated SB does not use random numbers, unlike SA, and thus is as deterministic and simple for parallel implementation as the original SBs (aSB, bSB and dSB).

A larger Ising machine can, in principle, be built by partitioning a spin system into multiple subsystems. In this case, the spin-spin couplings over the subsystems must be incorporated, and the partitioned subsystems also have to evolve in a single time domain. Communication and synchronization between the partitioned subsystems can easily degrade the speed performance. Tatsumura *et al.* proposed and demonstrated a scale-out architecture for SB-based Ising machines that enables continued scaling of both the machine size and speed performance by connecting multiple FPGAs as shown in Fig. 2 [5]. To maintain time consistency between multiple chips and a sufficiently small stall rate for every SB time step, the architecture relies on an autonomous synchronization mechanism that is implemented in the information exchange processes between neighbouring chips.

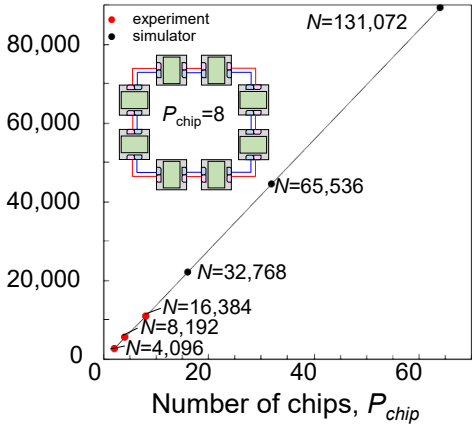

**Figure 2.** Scale-out architecture for SB [5]. Constant-efficiency scaling characteristic. (Inset) Connection of multiple chips in a bidirectional ring topology.

As an example of the application of SB accelerators for real-time systems, Tatsumura *et al.* presented an ultrafast financial transaction machine with a total response time of about 30 microseconds, including not only the detection of the most profitable cross-currency arbitrage opportunities by SB but also issuing order packets [7]. The detection problem of currency arbitrage opportunity was reduced to an optimal path search in a directed graph called a market graph, further formulated as an Ising problem, then solved with an SB accelerator. Steinhauer *et al.* used SB, in the financial field, for solving the integer portfolio and trading trajectory problem [8]. Zhang *et al.* applied bSB to traveling salesman problems and reported better solution accuracy and higher speed than an SA implementation [9]. Matsumoto *et al.* presented a hybrid method that iteratively uses a general-purpose processor (CPU) and an SB-based Ising machine for solving a discrete optimization problem (a distance-based clustering) with a complicated cost function (fractional-type) [10]. The complicated discrete problem is reformulated to an iterative algorithm including a step that solves an Ising problem. To minimize the communication overhead between the CPU and Ising machine, a low-latency implementation of SB was realized.

**Concluding Remarks**
Simulated bifurcation is a recently proposed, quantum-inspired, and highly parallelizable algorithm for combinatorial optimization. The high parallelism with massively parallel implementation technologies leads to high speed and scalability. The FPGA-based and GPU-based SB machines have been competitive against other cutting-edge Ising machines and have shown the highest performance for Ising problems with full connectivity. Massively parallel implementations of SB need many multiply-accumulators, large-capacity on-chip memory, and low-latency communication interfaces, and would best benefit from emerging nanodevices in conjunction with in-memory computing, stochastic computing, and cluster computing architectures. Integrating SB accelerators with other system components on a processing chip enables combinatorial optimization in real-time systems, and will offer new innovative applications.

## 8.2– Computing with Ising Machines realized through coupled nano-oscillators

Vito Puliafito, Politecnico di Bari, 70125 Bari, Italy (vito.puliafito@poliba.it)
Johan Åkerman, University of Gothenburg, 41296 Göteborg, Sweden (johan.akerman@physics.gu.se)
Hiroki Takesue, NTT Corporation, Atsugi, Kanagawa, 243-0198 Japan (hiroki.takesue@ntt.com)

**Status**

The solution of Combinatorial Optimization Problems (COPs) is currently of great interest for industrial applications, especially considering problems which are NP-hard, and their complexity scales exponentially with the number of the variables defining the phase space.

Ising Machines (IMs) are hardware solutions for the minimization of the cost function defined by the Ising model. This model describes the dynamics of $N$ spins $\sigma$ ($\sigma \in \pm 1$) through the following Hamiltonian:

$$IH = -\sum_{i,j=1}^{N} J_{i,j}\sigma_i\sigma_j - \sum_{i=1}^{N} h_i\sigma_i \, , \tag{1}$$

where $J$ is the matrix of coupling among the spins and $h$ is a local bias field. This research field is important because the minimization of $IH$ is NP-hard and several COPs with direct impact in logistics, manufacturing, financial management and artificial intelligence can be mapped into Ising model [1].

Several physical approaches have been used for implementing IMs and can be broadly divided into two categories: annealers and dynamical solvers. The former are physical systems that can reach the minimization of their energy (corresponding to the minimization of $IH$) by means of a gradual decrease of the temperature, through different thermal equilibrium states. They have been realized with optical systems, magnetic devices, memristors, CMOS circuits, and FPGAs, to cite a few. Dynamical solvers are characterized by a temperature-independent evolution towards the minimization state where a supplementary annealing process can speed up the process. They are mostly based on the coupling of oscillators, giving rise to the so-called Oscillator-based IMs (OIMs), and implementations include analog electronic [2], integrated CMOS [3], $VO_2$-based [4], spintronic [5], [6], spin waves [7] and optical coupled oscillators [8], [9] (see Fig. 1 for a few examples).

The most important aspect of OIMs is their scalability, which can guarantee the possibility to solve COPs with large number of spins densely connected. The investigation of this aspect can be easily performed through software solvers for the corresponding Ising models, which are of great support for a practical use of hardware IMs on the market.

Here, we concentrate on the two more promising and unconventional solutions of OIMs, those based on spintronic and optical oscillators. From a theoretical point of view, those machines can be simulated in software by using the well-established Kuramoto model. Calculations show a great potential in creating arrays as large as million nodes.

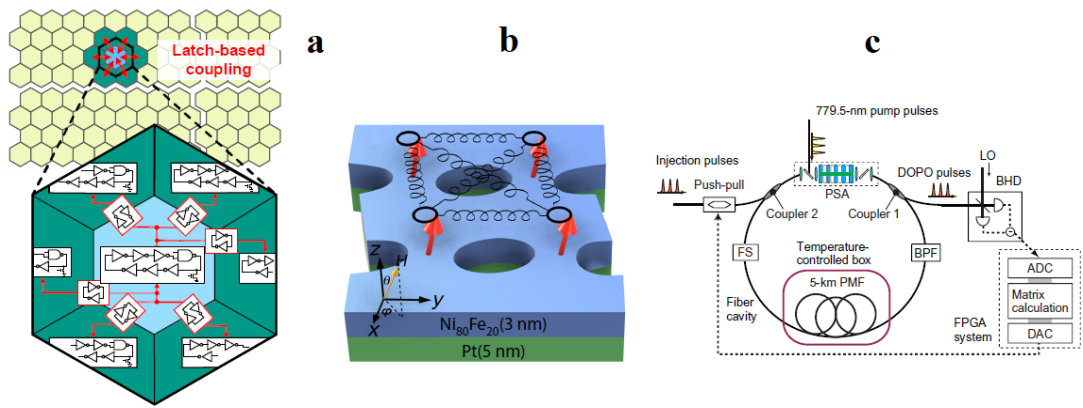

**Figure 1.** Sketches of different implementations for Ising Machines realized through coupled oscillators of different types: a) Hexagonal ring-oscillators [3] b) nanoconstriction-based spin-Hall nano-oscillators [6], c) optical parametric oscillators [9].

**Current and Future Challenges**

IMs have been studied and tested to challenge the most important limits of conventional computing, such as computational time, scalability and high integration, and possibility to approach the optimal solution of a large size COP, with a particular reference to the Max Cut Problem (MCP).

Spin-torque and spin Hall nano-oscillators (STNOs and SHNOs) have an attractive combination of properties, such as easy tunability, GHz frequency operation, and nanoscale size. They have been proposed for realizing OIMs in a theoretical approach making use of a universal model for a non-linear oscillator where sub-harmonic injection locking (SHIL) is implemented [5]. More recently, an experimental demonstration of a 2x2 array of nano-constriction SHNOs was realized showing binarization of their phases (fig. 1c) [6]. In the former study, the probability to solve a MCP remains very close to 98% up to about 180 nodes in Mobius graph, whereas in the latter, an estimation of 5000 SHNOs highlighted better properties with respect to a reference quantum solution. Spintronic oscillators, therefore, are very promising, but practical large-scale coupling between them requires additional development if complete and programmable all-to-all connections are required.

The use of degenerate optical parametric oscillators is the key-point of coherent IMs (CIMs), where each spin is encoded in the phase of light in an optical mode and oscillators are either in-phase or out-of-phase with respect to pump light (fig. 1d) [8], [9]. Spin connections can be realized through a network of optical delay lines, but a solution for a fully programmable all-to-all connections has been realized through an architecture that uses measurement-feedback [8]. In this case, a Mobius graph of 100 nodes has been used to test the MCP with a 21% of success probability, and fair solutions have been obtained for 2000-node MCPs [8]. CIMs have been compared to D-wave quantum annealers showing a more efficient performance in case of dense COPs. More recently, the MCP for a huge number of 100000-node graph has been solved with a CIM providing very good solution, comparable with those obtained by standard algorithms and annealers, in a shorter time to solution [9].

Software approaches take advantage of analytical models for the oscillators used in OIMs to predict their properties. The most famous model of mutually coupled oscillators was defined by Kuramoto [2]. It predicts that stability occurs when the phase difference between the oscillators is 0 or $\pi$, which can be obtained through an external signal at double frequency, introducing what was later called SHIL. The model has been developed and tested for solving COPs as well as for image processing, and it has been used as a reference for realizing physical implementations [2]. It has been tested for problem sizes ranging from 800 to 3000. Recently, an OIM based on a model for oscillators with frequency-phase coupling has been simulated obtaining the solution of the largest-size MCP so far, in a 2 million-node cubic graph [10]. Challenges include the increase of the graph connectivity and additive annealing techniques to speed up the time-to-solution.

**Advances in Science and Technology to Meet Challenges**
SHNOs have been demonstrated down to 20 nm, can operate at about 100 uA of current and 26 GHz, have been mutually synchronized in two-dimensional arrays of up to 64 oscillators, and individual SHNOs and their coupling to nearest neighbours can be controlled by voltage gates. While all these numbers need further improvement, the most fundamental limitation is the planar topology of nearest-neighbour coupling, which must be overcome. However, it can be shown that if next-nearest neighbour interactions are included, with or without control, e.g. along the diagonal in square arrays, this results in a non-planar topology. Fundamental research and experimental demonstrations in this direction are therefore needed.
CIMs seem to have a great potentiality to solve graphs of larger and larger number of nodes. Here, an important future challenge is to clarify how quantum nature of degenerate optical parametric oscillators contributes to the computational performance of CIMs.
The optimization of annealing techniques will be the advancement required not only for hardware IMs but also for algorithms. The latter ones will take advantage also from parallelization methods, while the inclusion of phase-power coupling in the model should be investigated to optimize the time to solution.

**Concluding Remarks**
IMs are on the crest of the wave nowadays, and they will surf it for the next years. Many solutions are on the table, and it is not trivial to compare them due to the several requested properties, such as a large number of nodes and a short time to solution, to cite the most important ones. In this scenario, IMs based on coupled oscillators guarantee advancements in technology and wide research activity in the upcoming future.

**Acknowledgements**
This work was also supported by Project No. PRIN 2020LWPKH7 funded by the Italian Ministry of University and Research, the Swedish Research Council Framework Grant no. 2016-05980, and the Horizon 2020 research and innovation programme (ERC Advanced Grant No.~835068 "TOPSPIN").

# 5.4– MemComputing: an opportunity for nanotechnology

Massimiliano Di Ventra, Department of Physics, University of California, San Diego, La Jolla, CA 92093, USA (diventra@physics.ucsd.edu)

Yuriy V. Pershin, Department of Physics and Astronomy, University of South Carolina, Columbia, South Carolina 29208, USA (pershin@physics.sc.edu)

**Status**

Any useful computing technology must satisfy one important goal: to aid in the computation of problems that are particularly challenging for us, the users. It is with this goal in mind that MemComputing was first suggested [1].

MemComputing is a new computing paradigm in which *time non-locality* (memory) and massive parallelism play the main role in the processing of information [2]. Time non-locality is the ability of a physical system to remember its past dynamics. The machine can then exploit it to solve the necessary tasks. The concept is radically different from the way our traditional computers, based on the Turing paradigm of computation, operate and even how quantum computers manipulate information. In addition, time non-locality is a feature shared by both quantum and non-quantum dynamical systems. This is not a minor point, since non-quantum dynamical systems offer substantial advantages for computing compared to quantum systems, in terms of both fabrication and simulation.

MemComputing machines have been mathematically defined in [3], where it was formally shown that they are Turing-complete, namely any problem solved by a Turing machine can be solved by a MemComputing one. In addition, it was shown that a MemComputing machine with specific features can solve NP-complete problems in polynomial time [4]. This result then begs the question: is this just a theoretical, albeit interesting outcome of their mathematical definition, or such a machine can be actually built in hardware? To answer the above question, a practical realization of *digital* MemComputing machines (DMMs) was proposed [4], which maps finite strings of symbols, such as 0 and 1, into a finite string of symbols, and relies on a new type of `self-organizing gates' (SOGs), namely gates which satisfy their logical truth table or algebraic relation irrespective of whether the signal is fed to the traditional input or output terminals (Figure 1).

Up to now, DMMs have been only simulated. Some of the results include the efficient solution of several optimization problems [2], acceleration of deep learning [5], and efficient solution of Boolean satisfiability problems [6], see Figure 2. These results were obtained by simply simulating the ordinary differential equations of DMMs [2]. In fact, for certain types of industrial problems such an `off-line' solution is sufficient [7]. These results then show that a physics-based approach to computation, like MemComputing, offers advantages not easily achievable by traditional algorithms.

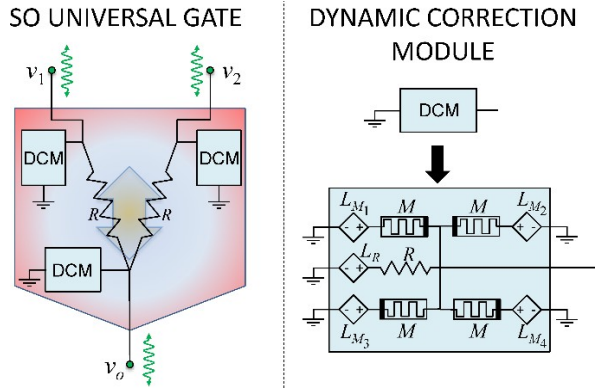

**Figure 1.** Left: Schematic of a self-organizing gate: the gate attempts to satisfy its logical proposition irrespective of whether the signal comes from the traditional input or output. This is possible thanks to the dynamics of the internal state variables employed in DCM modules (right). From [7].

**Current and Future Challenges**
However, for certain types of problems that are prominent in, e.g., autonomous vehicles, robotics, cryptography, and so on, a `real-time' solution is desirable. In turn, this requires a hardware realization. It is then the practical, hardware realization of these SOGs of MemComputing machines which is emerging as an important research direction, and we believe, will be a major focus of future research as well.

For instance, in [4] resistive memories and active elements were suggested as a way to implement SOGs and circuits built out of them. Since resistive memories can be emulated using complementary metal-oxide semiconductor (CMOS) technology [8], a full CMOS implementation of these machines is doable. In fact, since the size of problems that are relevant to industry and academia can easily reach millions of variables and constraints, a CMOS-based implementation seems the most reasonable first step towards hardware. Such realizations could be based either on field-programmable gate arrays (FPGAs) or application-specific integrated circuits (ASICs). We expect that these technologies will deliver a real-time MemComputing solution for some relevant industrial problems (providing at least 10x-100x speedup).

However, CMOS may not be ideal for low-power applications. In view of this, in [9], nanomagnetic SOGs have been suggested. In particular, a NAND gate (which is functionally complete) has been proposed that employs two main properties. First, by appropriately tailoring stray-field interactions between magnetized nanomagnetic islands one can enforce the logic proposition of the gate with equal population of all correct states. Second, a local dynamic error suppression scheme can be applied to limit the time spent in excursions between logically correct states, as a result of thermal fluctuations.

Another interesting research direction is to use spintronic resistive memories to build SOGs and their circuits [10]. For instance, magnetic tunnel junctions, controlled by an electric current or a magnetic field, can be employed as a low-power realization of these gates. In addition, since antiferromagnets have been shown to support resistive memory features, one can envision their use in MemComputing for very fast (THz range) operations.

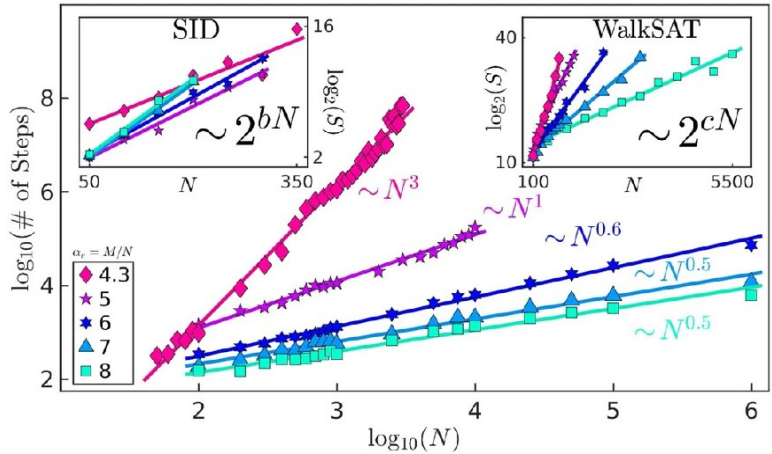

Figure 2. Polynomial scalability of time to solution of 3-SAT instances at fixed clause-to-variable ratio found by simulating DMMs. Insets: exponential scalability of classical algorithms (stochastic local-search algorithm, WalkSAT, and a survey-inspired decimation procedure, SID) on the same instances. From [10].

**Advances in Science and Technology to Meet Challenges**

The technological advances needed to realize the MemComputing paradigm in hardware are defined by various factors such as the selected technological platform, type of applications these machines will be used for, and the computing environment they will need to operate in. For instance, the CMOS and hybrid realizations require advancements in circuit theory that would allow the efficient implementation of the differential equations in terms of binary electronic circuits.

Moreover, the CMOS realizations of DMMs will require the development of a different theory: the theory of MemComputing maps. As the changes of states in digital electronic circuits are discrete (at times defined, e.g., by the clock cycle or cycles), the evolution of binary MemComputing circuits is a map. In this case, particular care needs to be given to two important aspects of maps that may appear in the transition from continuous dynamical systems [2]: *i*) maps may introduce extra critical points in addition to the ones of the original dynamical system. These are called ghost critical points.

*ii*) The basin of attraction of the equilibrium points may shrink for maps. This direction of study is then very important.

Regarding the type of applications of these machines, we need to stress that typical problem instances of interest in both academia and industry involve hundreds of thousands or even millions of variables and constraints. Such problems then would require a level of integration that is not easily achievable outside of CMOS technology. Therefore, emerging realizations of SOGs and circuits built out of them using nanotechnology components must also satisfy the high bar of being scalable and possibly compatible with CMOS.

Spintronics seems to have many features that would allow such a hybrid CMOS-based realization of MemComputing [10]. In fact, magnetic tunnel junctions, one of the basic ingredients of spintronics, can now be integrated with CMOS .

**Concluding Remarks**

We have started this article with an important point worth repeating. The usefulness of *any* technology should be first addressed in terms of its end goal. In the case of computing, the goal is to solve problems that are challenging for us, the users. If a computing machine does *not* accomplish such a goal, even if academically interesting, it is not practically useful [2]. In this respect, MemComputing has already shown several advantages compared to our conventional computing model and other paradigms of current interest, such as quantum computing. These advantages reflect first in the possibility of emulating DMMs in software, thus allowing a direct comparison with traditional algorithms. Second, these machines do not rely on quantum phenomena, like entanglement, to work. Therefore, the path towards hardware is considerably less challenging than for quantum computers. In fact, DMMs can even be realized using our standard CMOS technology, providing an opportunity for very-large-scale integration. Finally, we expect MemComputing, or any other `unconventional computing' paradigm, to extend the reach, and enhance the functions of our modern computational fabric, not to replace it. In other words, we expect MemComputing machines to play the role of *co-processors* specialized to tackle particularly challenging problems.

**Acknowledgements**
M.D. and Y.V.P. acknowledge support from the National Science Foundation under Grant No. ECCS-2229880.

# 6.1 – Compute-in-Memory with Nanoscale CMOS Technologies

Amit Ranjan Trivedi, Department of Electrical and Computer Engineering, University of Illinois at Chicago, Chicago, IL 60607, USA (amitrt@uic.edu)

Saibal Mukhopadhyay, School of Electrical & Computer Engineering, Georgia Institute of Technology, Atlanta, GA 30332, USA (saibal.mukhopadhyay@ece.gatech.edu)

Kaushik Roy, Elmore Family School of Electrical and Computer Engineering, Purdue University, West Lafayette, IN 47907, USA (kaushik@ecn.purdue.edu)

**Status**

Deep learning algorithms have shown that the growing volume and variety of data can be leveraged for highly accurate predictions and decision-making in many complex problems. A deep neural network (DNN) typically utilizes millions to billions of weights (i.e., model parameters). Operating over such a large parametric space facilitates the network with robust inductive biases. Yet, it also presents critical inference constraints for real-time or low power applications. Especially, DNN's extensive model size induces excessive memory accesses to read model weights from off-chip memories and to read/write operands to off-chip memory and intermediate memory hierarchy. Thus, on a conventional digital hardware, the inference performance of DNN succumbs to limited processor-memory bandwidth. A radical approach gaining attention to address the performance challenge is to design alternate non-von Neumann computing modules that can not only store model weights but also locally process most inference operations within the same structure. Therefore, using such "compute-in-memory" processing of DNN, high volume data traffic between processor and memory units can be averted. Compute- in-memory using conventional CMOS-based memory structures is especially more promising. Prior works have shown that CMOS-based conventional memory structures such as SRAM, DRAM, embedded-DRAM, SONOS, and NAND-Flash, *etc.*, can be adapted for compute-in-memory, thus enabling a rapid and cost-effective adoption of the scheme in commercial substrates.

Matrix-Vector multiplications (MVM) constitute the dominant computations in a DNN. To leverage CMOS memories for the storage of model weights and MVM computations, most compute-in-memory schemes employ a mixed-signal processing. Digital inputs to a DNN layer are converted to analog representation such as charge [1], current [2], or time [3]. The input vectors are loaded in parallel to the memory array where the memory cells multiply them with the stored weights in an analog fashion. The analog output of all memory cells within a column is summed to produce the output of MVM. Especially, the accumulation of products in many schemes simply reduces to current/charge summation over a wire, thus further minimizing the necessary workload. The analog MVM outputs are subsequently digitized for storage and routing to the other processing units.

Among early works on CMOS-based compute-in-memory, Zhang et. al. presented the processing using an array of standard six-transistor (6T) SRAM cells [2]. The resulting memory array, however, was vulnerable to instability under process variability. The challenges were resolved in [1] using 10T SRAM cells which separated ports for inference and write. While early adoptions of CMOS-based compute-in-memory focused on binary weights, the schemes were later enhanced for multibit processing to improve the accuracy of DNN. Detailed survey of compute-in-memory processors was compiled in [4].

**Current and Future Challenges**

Most compute-in-memory schemes employ mixed-signal processing which raises critical challenges to integrate analog circuits such as analog-to-digital converter (ADC), digital-to-analog converter (DAC), and comparator within the memory structures. For example, in CONV-SRAM [1], to compute the inner product of $l$-element weight and input vectors **w** and **x**, $l$-DACs and one ADC are required. Since DACs are concurrently active, they lead to both high area and power. Since most memory modules are designed using advanced nanometer node CMOS technologies for energy and area efficiency, designing memory-integrated analog circuits at the same technology node is challenging. At such advanced technology nodes, the analog circuits are susceptible to failure under process variability and require complicated calibration processes. Due to such processing challenges of mixed-signal operations in computing-in-memory, significant research in the past many years has focused on exploring alternatives to alleviate the implementation complexity, especially at advanced CMOS nodes.

In [5], time-domain DACs were used to replace analog circuits for DAC implementation. However, with increasing input precision, either operating time increases exponentially, or complex analog-domain voltage scaling is necessitated. All digital compute-in-memory processing with SRAM was shown for binary neural networks, e.g., in [6]. However, for more complex deep learning applications such as object detection and autonomous navigation, networks with binary-weighted inputs and weights have very low accuracy. The accuracy of SRAM-implemented binary networks was improved using supported-BinaryNet architecture in [7] and by leveraging peripherals DACs to implement the support parameters. Implementation of multiplication-free neural networks was discussed in [8] by adapting deep learning's inference operator such that the multiplications between multibit precision weight and input vectors were not necessary. Novel adaptations of SRAM-based compute-in-memory were discussed where the memory structure was employed for Monte Carlo Dropout (MC-Dropout) in [9]. Recently, neural network transformations to frequency-domain were exploited to design ADC/DAC-free analog acceleration for multibit precision input/output in-memory processing while enabling perfect parallelism along input vector and output vector computations as shown in Figure 1 [10]. Such end-to-end co-design methods bridging the design of nanoscale compute-in-memory structures to mapped algorithm has shown a remarkable promise where the co-designed structures have shown orders of magnitude improvement over the traditional designs.

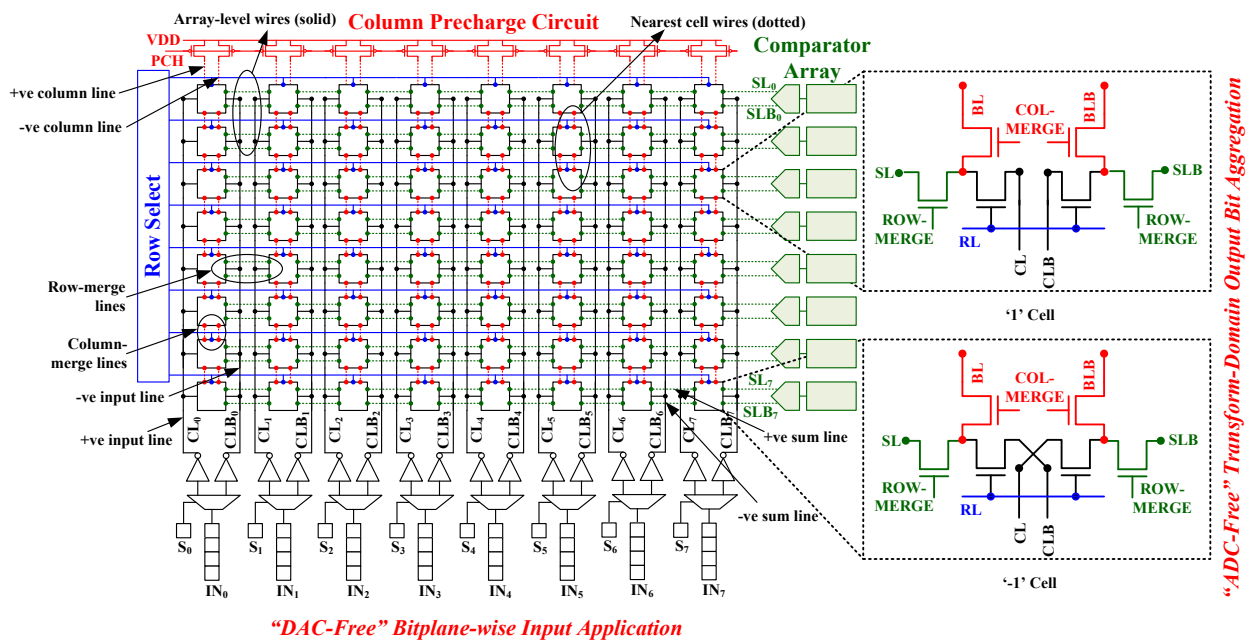

**Figure 1:** Compute-in-SRAM designs exploit analog computations to exploit "physics for computing," however suffer due to the use of many digital-to-analog converters (DAC) and analog-to-digital converters (ADC) for each memory array which limits their scalability and energy

efficiency. Novel implementation methods have exploited co-design methods such as transform-domain neural network operations to overcome these limitations [10]. The *co-designed approach* has achieved orders of magnitude more energy efficiency and significant scalability over the traditional compute-in-memory implementation.

**Advances in Science and Technology to Meet Challenges**

Integrating computations and storage invariably demands more area per cell in compute-in-memory. Meanwhile, state-of-the-art DNNs continue to increase in model-size, thereby demanding higher energy and area-efficiency of the memory structures. In the future, several complementary efforts must be pursued in cohesion to improve the area efficiency of compute-in-memory. Compute-in-memory inference architectures that can robustly operate in more advanced CMOS nodes, such as 7 nm or below, will be imperative. Compute-in-memory in monolithic and vertically integrated memory structures need to be pursued. Low and mixed precision DNNs, better suited for compute-in-memory processing, will be needed. Pruning and compression methods of DNN will be critical. Almost or completely digital architectures will be needed that maintaining multibit precision operations as well as the advantages of analog mode processing such as minimizing workload by exploiting physics for computations. In parallel, DNN architectures themselves are going through a dramatic evolution to improve their computational efficiency. In the last few years, novel layers such as inception, residual layers, dynamic gating, polynomial layers, self-attention, and Hypernetworks have been added to the repository of DNN building blocks. Therefore, a critical challenge for the next generation compute-in-memory accelerators for DNN is to exhibit high versatility in their processing flow for efficient mapping of these diverse DNN layers into hardware circuits. Especially, many emerging layers, unlike classical layers, simultaneously correlate multiple variables to enhance computational efficiency and representation capacity. Therefore, novel compute-in-memory schemes will be needed to map higher-order processing of the emerging layers within simplified cells.

## 6.2 – In-memory computing using non-volatile memories

I-Ting Wang, Taiwan Semiconductor Research Institute, Hsinchu 300091, Taiwan(itwang@narlabs.org.tw)

Wang Kang, School of Integrated Circuit Science and Engineering, Beihang University, Beijing 100191, China(wang.kang@buaa.edu.cn)

Yao Zhu, Institute of Microelectronics, Singapore 117685, Singapore (zhuya@ime.a-star.edu.sg)

Brajesh Kumar Kaushik, Department of Electronics and Communication Engineering, Indian Institute of Technology-Roorkee, Roorkee, Uttarakhand 247667, India (brajesh.kaushik@ece.iitr.ac.in)

**Status**

In-memory computing (IMC) using emerging non-volatile memories (NVMs) has successfully opened up new opportunities for future computing paradigm. The NVMs, including resistive random access memory (RRAM), ferroelectric RAM (FRAM), and magnetoresistive RAM (MRAM), resemble artificial synapses with adjustable conductance as synaptic weight. In particular, NVM-based synaptic device is provided with computing and storage functions simultaneously, which eliminates the inefficient data movement between physically separated processor and memory units in conventional von Neumann computing architectures (Fig. 1 a). Besides, NVM-based IMC provides massively parallel computation in a crossbar configuration to further boost the computing efficiency. Therefore, an enormous amount of research has been devoted to developing NVM-based synaptic devices for building a high energy- and area-efficient computing hardware.

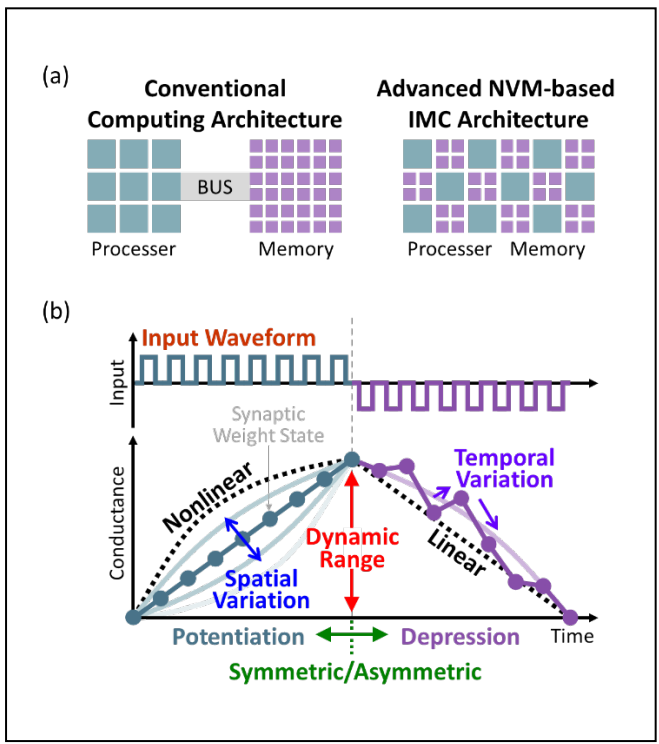

**Figure 1.** (a) Comparison between conventional computing architecture and NVM-based IMC architecture, where the former suffers from the inefficient data transfer between physically separated processing and memory units while the latter realizes computing and storage functions in the same location to boosts the computing efficiency. (b) General behaviors in the NVM-based synaptic device with bidirectionally adjustable synaptic weight states through the given input waveform. The non-ideal device properties such as nonlinear/asymmetric weight modulation and temporal/spatial variation are indicated accordingly.

RRAM with inherently two-terminal structure ensures a compact and high-density synaptic array that simply performs the multiplication of the input signal and synaptic weight state with high computing parallelism in the neural network. Therefore, fruitful results from the device-, circuit-, and system-level demonstrations have been extensively presented [1].

The revival of FRAM has rapidly attracted increasing attention since the unprecedented discovery of ferroelectricity in hafnium oxide (HfO$_2$) [2], where the ferroelectric HfO$_2$ successfully solved the limitations of complementary metal-oxide-semiconductor (CMOS) process incompatibility and scalability in conventional perovskite oxides. In particular, the HfO$_2$-based ferroelectric field-effect transistor (FeFET) not only promises fast operating speed and low energy consumption due to the field-driven domain switching, it also provides stable multi-state with partially switched domains in the ferroelectric. These superior properties make HfO$_2$-based FRAMs stand out from the currently developed synaptic devices.

MRAM utilizes the spin property of electrons and holds the promise of low power, high speed and high endurance IMC. Very recently, implementation of IMC artificial neural networks (ANNs) based on MRAM crossbar array has been demonstrated, offering a potential platform to mimic the brain [3]. After the commercialization of spin-transfer torque MRAM, recently, the spin-orbit torque MRAM and the voltage-controlled magnetic anisotropy MRAM as well as the combination of the two effects are under intensive investigations, targeting further reduction of power and latency.

**Current and Future Challenges**

Figure 1b summarizes the behaviors commonly found in the NVM-based synaptic devices. Generally speaking, an ideal synaptic device should have a bidirectional modulation (i.e., potentiation and depression) in synaptic states that fulfils the requirements of computing and storage. First, the synaptic states should be non-volatile and free from spatial variation among device-to-device. Second, the synaptic modulation should be linear and symmetric and without temporal variation from cycle-to-cycle. Besides, high endurance in synaptic modulation by identical input waveform is important not only to guarantee a sufficient training epochs but also prevent a time-consuming read-before-write process. Moreover, an ideal synaptic device should expand in an adequate dynamic range compatible with that of peripheral circuit because the higher conductance results in additional energy consumption while the lower increases sensing latency. However, even though a tremendous amount of work has been carried out for searching the holy grail, the truly ideal synaptic device is still lacking, which has left ample room for improvement and compromise.

Although RRAM-based synaptic device is relatively matured for realizing hardware neural network, it still inevitably suffers from non-ideal device properties, which significantly impact on accuracy degradation [4]. Improvements and optimizations from material/device engineering are therefore important. However, adopting novel materials such as two-dimensional transition metal dichalcogenides (2D TMDs) is still under scrutiny and thus lacks of statistic data to support its practicability. Moreover, selecting elements must be implemented to suppress the unwanted leakage current and interference from the unselected cells, but the process complexity is increased. Developing a self-selecting and self-rectifying synaptic device without the need of selecting element is still challenging.

As for the novel HfO$_2$-based FeFET, integrating the ferroelectric gate stack at font-end-of-line process compromises the write efficiency and performance. By adding an additional floating gate between ferroelectric and gate insulator, the *m*-MFMFET [5] not only solved the above-mentioned issues, but it also provided back-end-of-line (BEOL) fabrication flexibility to ease the hardware design. However, scaling down the ferroelectric in both vertical (thickness) and horizontal (cell dimension) directions under BEOL-compatible process temperature while maintaining sufficient remanent polarization is still challenging. Besides, although FeFET-based synaptic device may improve the variability issue due to the relatively stable spontaneous polarization, achieving linear synaptic modulation using identical input waveform is still challenging.

Regarding MRAM, despite its practical advantages in power, endurance and technology maturity, the difficulty for high-performance IMC hardware stems from the low absolute resistance (~several kOhm) and the low on/off resistance ratio (~300%) of MRAM devices, which bring challenges in implementing large-scale multi-bit computing, e.g., analogue multiply–accumulate operations. Techniques from devices/circuits co-engineering to design new computing paradigms or architectures are therefore important.

**Advances in Science and Technology to Meet Challenges**
Although IMC with NVMs is promising for future computing paradigm, it still remains rooms for improvement. Several possible directions that we anticipate are described as follows:

First, continued optimization and innovation in material/device engineering are the keys, where the inevitably intrinsic variation and non-ideal device properties could be greatly improved. For instance, 2D TMDs with ferroelectricity are recently found and reported with promising scalability and reliability [6], which may shed some light on the hardware neural network. Besides, the recent discovery of aluminum scandium nitride (AlScN) with superior ferroelectric properties such as high remanent polarization and tightly distributed coercive field may become another relevant candidate [7]. Moreover, alternative device structures for different computing paradigms need to be explored. Skyrmionic MRAM devices for reservoir computing and probabilistic/stochastic computing are good examples to further exploit the device features [8].

Meanwhile, a neural network evaluating platform with holistic optimizations from device-, circuit-, and system-level for ANN design guideline and performance prediction is especially crucial to continuously take pre-emptive actions for constructing future computing hardware. A general standard for measuring the synaptic device is usually based on that for the NVMs [9]. However, their characteristics for application-specific criteria are found to be much relaxed for ANN applications [10]. A more practical evaluating methods suitable for the synaptic devices is thus required.

Finally, by leveraging the strengths in the NVMs and the matured CMOS components, developing novel computing architecture with both IMC and digital hardware designs may further relive the device requirements since the non-ideal properties in the NVM-based synaptic device unavoidably exist. By pulling together different devices with their superiorities, the hybrid architecture promises the best trade-off that is surely worth developing.

**Concluding Remarks**
In-memory computing with the emerging NVMs is gaining great momentum in research as the data-centric tasks no longer be affordable in conventional computing architectures. By taking advantage of the NVM crossbar array, the NVM-based IMC is promising for massively parallel computation, which successfully improves the computing efficiency. However, each NVM device has its own issues that are mostly attributed to the intrinsic and non-ideal device properties, and it therefore remains rooms for improvement. It is worth mentioning that to pursue a more practical IMC hardware, investigation involving with device, circuit, and system co-optimization is more desirable compare to that of merely focusing on a single angle. Therefore, with these driving forces for resolving current challenges in multiple aspects, we anticipate that NVM-based IMC to be more energy- and area-efficient and continuously pave the way for leading-edge computing paradigm.

**Acknowledgements**
This work was partially supported by the Science and Engineering Research Council of A*STAR (Agency for Science, Technology and Research) Singapore, under Grant No. A20G9b0135.

# 7.1– Brain-Inspired Cortical—Hipocampus Inference and Lifelong Learning

Jennifer Hasler, School of Electrical & Computer Engineering, Georgia Institute of Technology, Atlanta, GA 30332, USA (Jennifer.hasler@ece.gatech.edu)

Samiran Ganguly, Department of Electrical and Computer Engineering, Virginia Commonwealth University, Richmond, VA 23284, USA (gangulys2@vcu.edu)

Avik W. Ghosh, Department of Electrical and Computer Engineering, University of Virginia, Charlottesville, VA 22904, USA (ag7rq@virginia.edu)

William Levy, Department of Electrical and Computer Engineering, University of Virginia, Charlottesville, VA 22904, USA (wbl@virginia.edu)

Vwani Roychowdhury, Department of Electrical and Computer Engineering, University of California at Los Angeles, CA 90095, USA (vwani@ee.ucla.edu)

Supriyo Bandyopadhyay, Department of Electrical and Computer Engineering, Virginia Commonwealth University, Richmond, VA 23284, USA (sbandy@vcu.edu)

**Status**
In the world of exploding data, there is an urgent need for brain-inspired computing that goes beyond conventional artificial neural networks. Today's ChatGPT or DALL-E are enabled through large clusters of GPU/TPU/FPGA architectures, requiring huge levels of energy for training these networks [1]. Conventional machine learning (ML) based on Deep Neural Nets (DNN) is ill-suited for the rapid growth in edge intelligence due to the considerable cost of wiring and a limit on memory resources. Edge intelligent machines must analyze rapidly varying situations with very limited energy and memory resources without relying on a cloud that is either unavailable or unreliable in the face of security breaches, such as autonomous robots exploring mines, battlefield vehicles navigating enemy terrain, or Mars rovers. For example, 3D Autonomous Simultaneous Localization and Mapping (ASLAM) for mobile robotics requires offline calculations for robot actions, traditionally done by deleting out-of-date data using a delayed nearest neighbor data association strategy [2-4]. Going beyond such a local adaptive mapping and navigation, one step at a time, becomes completely prohibitive over an expanded time horizon.

New ML sensory data are learned *independent* of past history. The resulting 'catastrophic forgetfulness' (sequential data over-write) requires added hardware and costly synaptic interconnects to deal with, with an exploding area footprint and energy budget. The challenge with learning sequential data is the *stability-plasticity dilemma*, the choice between integrating new knowledge vs remembering previous knowledge. Reducing overlap among stored internal representations, such as sparse or interleaved learning partially addresses the dilemma [5]. Connection networks absorb new inputs and adjust synaptic weights incrementally, thereby preventing sequential learning. Intel's recent robot [6] relies on a learning phase where prototype data are moved around but not erased in feature space, punishing the wrong category or boosting the right category, and only allocating new resources if the error persists and an unknown category is thereby identified.

Human cognitive abilities provide a contrasting example where the brain encodes continuous learning very efficiently with a fixed memory bank, avoiding catastrophic forgetfulness with incremental (and contextual) learning. The evolved trick is an over-riding hippocampus (HC) that acts as a memory support structure for a severely resource-constrained (with dense local and otherwise sparse, connectivity [7]) neocortex (NC) that mediates perceptions and decisions. In contrast to brain development which proceeds from sensory, up through the cortical hierarchy (i.e., bottom-up), later learning proceeds in a top-down manner directed by the hippocampal system, a dense recurrently connected generative network with the ability to process and correlate long term multi-dimensional sequences [8,9] to build a sophisticated "World Model" for the organism [Fig. 1]. In addition to the hardware support structure of a dual memory, the brain also uses algorithmic techniques to encode episodic memory by hierarchically implementing depths of representation. Big ideas are built out of association between contexts, event histories and multi-sensory microscopic details; for instance, reframing a 'zebra' as a 'striped horse'. Holiday might be contextualized with the sight of an apple pie and the smell of turkey in the oven; subsequently songs by Bing Crosby would refine that knowledge down to Christmas [Fig. 2]. *The process of having a separate dense hippocampal network directing a sparser neocortical network allows us to avoid catastrophic forgetfulness and compose coherent long-term semantic memory (lifelong learning) from short, fragmented, conditionally independent, episodic events and their intermediate representations.*

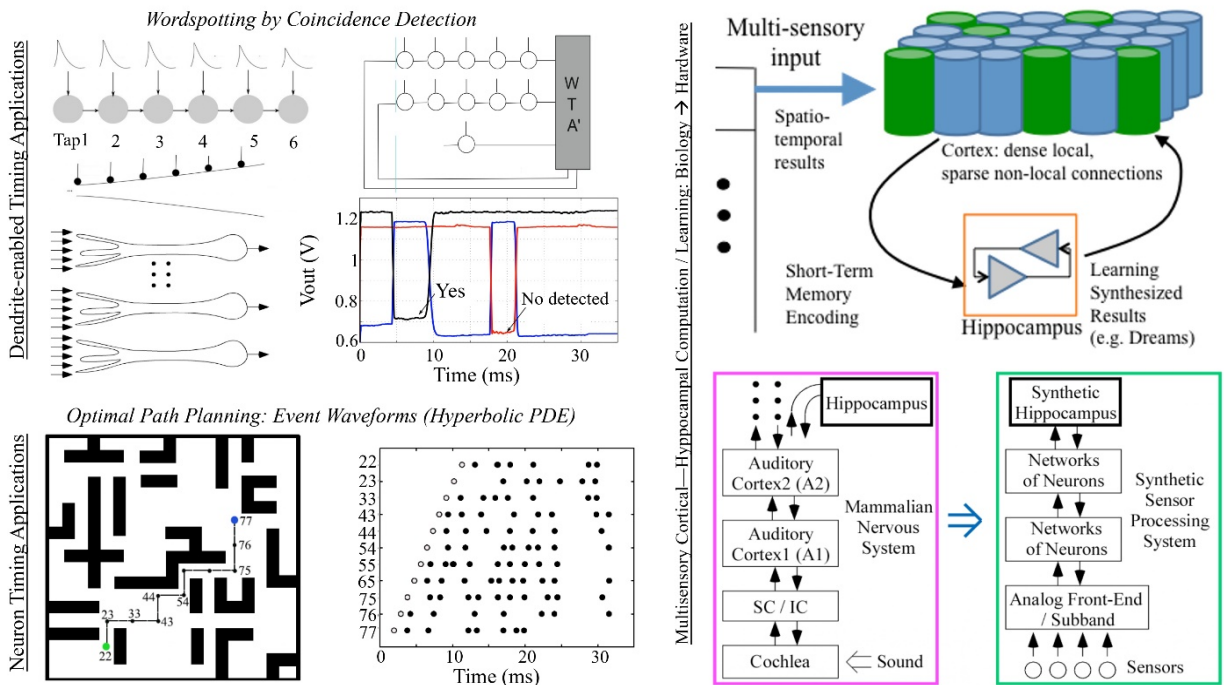

**Figure 1:** Neural computation requires spatio-temporal processing involving many modes of processing neuron output events that includes event-timing between neurons (e.g., optimal path planning) as well as event-timing (e.g., coincidence detection) within neuron dendrite (e.g., wordspotting). Timing is utilized throughout the neural infrastructure and sensory inputs, between the bidirectional computation of cortical layers, as well as with hippocampus. The cortex--hippocampus architecture performs multi-sensory data fusion by constructing meaningful semantics from episodic memories. Integrating this entire computing and learning biological model into a synthetic system becomes the challenge for next-generation energy-efficient neuromorphic systems.

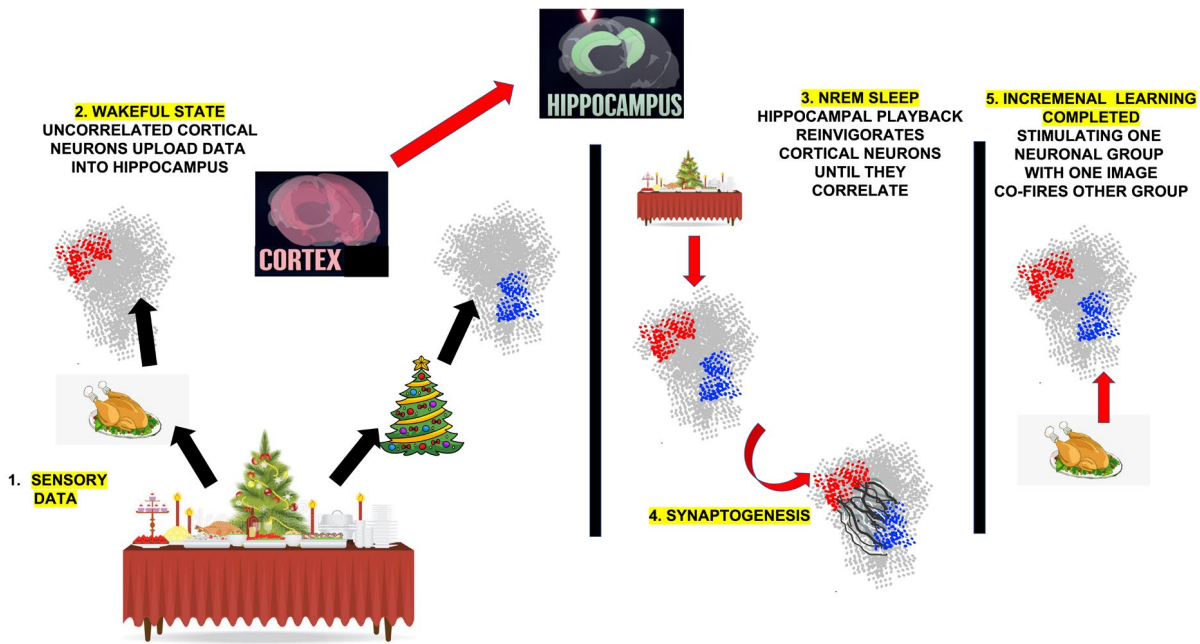

**Figure 2.** (Left) The uncoordinated uploading of sensory data from neocortex to hippocampus during wakeful state, is followed by a memory rehash through ~10Hz theta cycles (compressed time stamps), each consisting of ~40Hz gamma oscillations (synchronous events), during NREM sleep, allowing the relevant cortical neurons time to connect afresh (synaptogenesis) through Hebbian overlap.

**Current and Future Challenges**
The current and future challenge is to develop neural structures with these hippocampal—cortical inference and learning systems. Learning and computation requires using neural architectures that might include input sensors refining their input data [Fig. 1], processing through bidirectional layers of neurons, and a hippocampus layer (and related layers) for encoding and training. During awake activity, sensory signals are processed through subcortical layers in the cortex and the refined outputs reach the hippocampus. During the sleep cycle, these memory events are replayed to the neocortex where sensory signals cannot disrupt the playback. During sleep, hippocampal interactions strengthen the memory representations in the neocortex by strengthening some synapses and even establishing new synapses.

Building energy efficient brain-like neural systems requires rethinking models of neural computation utilizing the timing of neural events rather than simply encoding values in number of spikes, not just in single neurons, but in networks of thousands of neuron components [Fig. 1]. Unlocking these opportunities requires the efficient use of event-timing and temporal neural encoding as well as modeling and computationally abstracting the fundamental computations of 100s to 10,000s of cortical neurons (e.g., neural columns) [7]. These components organize into bidirectional interconnected cortical layers interacting with other neural layers (e.g., hippocampus) for computation and learning, where the learning process requires both parameter updates (e.g., synaptic weights) as well as new topologies and components (e.g., neurogenesis, synaptogenesis, dendritogenesis).

At the architecture and algorithm level, adaptive synaptogenesis [Fig. 2] relies on minimizing expensive distal synaptic connections by making them rare, driven by a flagging depth of neuronal activity and directed by an overseeing hippocampus. Through a replay of time-compressed and context-driven reconstruction of sensory experience, the hippocampus directs lower-level neocortical neurons into the appropriate Hebbian adjustment of their synaptic weights through 'wire-on-the-fly' generation and annihilation of distal connectivities, thereby allowing continual learning and eliminating catastrophic forgetfulness. Replicating this process of *synaptogenesis* with efficient hardware is an important future challenge.

**Advances in Science and Technology to Meet Challenges**
Building efficient event-timing networks becomes the primary challenge for synthetic neuromorphic systems. Extending computation and learning to an energy constrained [1] edge environment, with acceptable Size Weight and Power (SWaP) is the desired goal [10]. The original neural roadmap to develop a human cortex and a human brain (2013) showed a potential solution using Si CMOS technology utilizing the close connection between biological and Si devices [7]. Event-timing networks will require additional techniques to handle the range of timescales and morphological changes. Neurobiological systems operate over many orders of magnitude in timing and the learning on that timing and utilize structures like glial (and other) cells for the timing modulation. To date, many of these neuromorphic techniques using efficient temporal encoding of events have many untapped engineering application opportunities.

One example uses event-timing for predicting optimal paths through an array of neurons, where an optimal path is found by the first arriving events in a polynomial resource algorithm [Fig. 1]. Another example uses coincidence detection of event timing in efficient dendritic processes between a cluster of dendritic-enabled neurons [Fig. 1]. These techniques require physical computation to efficiently model the ordinary differential equations (ODEs) as well as the dendritic partial differential equations (PDEs) of cortical neurons (e.g., pyramidal cells) that includes the large dendritic arborization and as well as networks (e.g., cortical columns) of these pyramidal cells [7]. Both of these techniques result in significantly higher energy efficient computations compared to analog operations.

Several emerging nanotechnologies may accelerate and reduce energy cost of brain-inspired unconventional computation. Algorithmic advances and new device ideas incorporating perhaps non-traditional devices that are extremely frugal in their use of energy and can communicate with each other without wires (e.g., dipole coupled nanomagnets) can dramatically increase both SWaP and speed of computation. For instance, an array of nanomagnets can act as neurons [11,12], with inter-magnet dipole coupling acting as synapses, reducing energy consumption in inter-neuron communication dramatically since dipole coupling is "wireless" and current free. There is no conduction current flow and displacement current flows only when the weights are changed with electrically generated stress. There is accordingly no RC delay or area penalty associated with charging wireless interconnects. This can result in dramatic improvement of SWaP (size, weight and power). The synaptic weights can be modulated by varying the effective strength of dipole coupling between two nanomagnets fabricated on a piezoelectric substrate with gates that apply a local potential to the region of the piezoelectric pinched between the nanomagnets. The potential generates local stress which modulates the energy barriers within the nanomagnets and thereby modulates the effect of dipole coupling [13].

**Concluding Remarks**
Brain-inspired computing greatly expands the reach of current computing approaches focused on deep neural networks, recurrent networks, and others that do not utilize the rich and highly energy efficient neural system computing capabilities. It provides a pathway to lifelong learning, avoidance of catastrophic forgetfulness, and the ability to compute in extremely resource constrained environments. Building synthetic physical brain-inspired computing to achieve these cortex-sized goals with drastically improved energy efficiency over current digital techniques [1] has a number of opportunities both using analog techniques in standard CMOS (e.g. [7,10]) as well as potentially through emerging nanotechnologies (nanomagnets, memristors and even nano-CMOS).